\documentclass[aps,prl,twocolumn,showpacs,superscriptaddress,groupedaddress]{revtex4-2}
\usepackage{graphicx}  
\usepackage{dcolumn}   
\usepackage{bm}        
\usepackage{amssymb}   
\usepackage{amsmath}
\usepackage{xcolor}
\usepackage{amsthm}
\usepackage{enumitem}
\newcommand{\be}{\begin{equation}}\newcommand{\ee}{\end{equation}}
\newcommand{\bea}{\begin{eqnarray}}\newcommand{\eea}{\end{eqnarray}}
\newcommand{\brr}{\begin{array}}\newcommand{\err}{\end{array}}
\newcommand{\bit}{\begin{itemize}}\newcommand{\eit}{\end{itemize}}
\newcommand{\ben}{\begin{enumerate}}\newcommand{\een}{\end{enumerate}}

\newcommand{\ba}{\begin{array}}
\newcommand{\ea}{\end{array}}

\definecolor{darkred}{rgb}{.8,0,0}

\definecolor{darkblue}{rgb}{0,0,.7}

\def\lf{\left}

\def\ri{\right}
\def\al{\alpha}\def\ga{\gamma}
\def\de{\delta}

\def\1{{_{1}}}\def\2{{_{2}}}

\def\noHe0{:\;\!\!\;\!\!:H_e(0):\;\!\!\;\!\!:}
\def\noHm0{:\;\!\!\;\!\!:H_\mu(0):\;\!\!\;\!\!:}

\def\lf{\left}

\def\ri{\right}

\def\al{\alpha}\def\ga{\gamma}
\def\de{\delta}

\def\1{{_{1}}}\def\2{{_{2}}}

\begin{document}

\title{Decoherence limit of quantum systems obeying generalized uncertainty principle:
new paradigm for Tsallis thermostatistics}


\author{Petr Jizba}
\email{p.jizba@fjfi.cvut.cz}
\affiliation{FNSPE,
Czech Technical University in Prague, B\v{r}ehov\'{a} 7, 115 19, Prague, Czech Republic}
\affiliation{ITP, Freie Universit\"{a}t Berlin, Arnimallee 14, D-14195 Berlin, Germany}

\author{Gaetano Lambiase}
\email{lambiase@sa.infn.it}
\affiliation{Dipartimento di Fisica, Universit\`a di Salerno, Via Giovanni Paolo II, 132 I-84084 Fisciano (SA), Italy}
\affiliation{INFN, Sezione di Napoli, Gruppo collegato di Salerno, Italy}

\author{Giuseppe Gaetano Luciano}
\email{gluciano@sa.infn.it}
\affiliation{Dipartimento di Fisica, Universit\`a di Salerno, Via Giovanni Paolo II, 132 I-84084 Fisciano (SA), Italy}
\affiliation{INFN, Sezione di Napoli, Gruppo collegato di Salerno, Italy}

\author{Luciano Petruzziello}
\email{lupetruzziello@unisa.it}
\affiliation{Dipartimento di Fisica, Universit\`a di Salerno, Via Giovanni Paolo II, 132 I-84084 Fisciano (SA), Italy}
\affiliation{INFN, Sezione di Napoli, Gruppo collegato di Salerno, Italy}
\affiliation{Dipartimento di Ingegneria, Universit\`a di Salerno, Via Giovanni Paolo II, 132 I-84084 Fisciano (SA), Italy}

\date{\today}
\def\be{\begin{equation}}
\def\ee{\end{equation}}
\def\al{\alpha}
\def\bea{\begin{eqnarray}}
\def\eea{\end{eqnarray}}

\begin{abstract}
The generalized uncertainty principle (GUP) is a phenomenological model whose purpose is to account for a minimal length scale (e.g., Planck scale or characteristic inverse-mass scale in effective quantum description) in quantum systems.  In this Letter, we study possible observational effects of GUP systems in their decoherence domain.  We first derive coherent states associated to GUP and unveil that in the momentum representation they coincide with Tsallis’ probability amplitudes, whose non-extensivity parameter $q$ monotonically increases with the GUP deformation parameter $\beta$. Secondly, for $\beta < 0$ (i.e., $q < 1$), we show that, due to Bekner--Babenko inequality, the GUP is fully equivalent to information-theoretic uncertainty relations based on Tsallis-entropy-power. Finally, we invoke the Maximal Entropy principle known from estimation theory to reveal connection between the quasi-classical (decoherence) limit of GUP-related quantum theory and non-extensive thermostatistics of Tsallis. This might provide an exciting paradigm in a range of fields from quantum theory to analog gravity.  For instance, in some quantum gravity theories, such as conformal gravity, aforementioned quasi-classical regime has relevant observational consequences. We discuss some of the implications.
\end{abstract}

\vskip -1.0 truecm

\maketitle

{\em {Introduction}.}~---~
%
There are indications from various studies such as String Theory, Loop Quantum Gravity,
Quantum Geometry or Doubly Special Relativity (DSR) theories, that the uncertainty relation between positions and momenta acquire corrections due to gravity effects and should be modified accordingly~\cite{string,string II,ppp,rovelli,Ashtekar,MS1,MS2}.
%
%
These modifications implement, in one way or another, the minimal length scale and/or
the maximum momentum. The ensuing modified uncertainty relations are known as the generalized uncertainty principles --- GUP's.
A paradigmatic form of GUP is the quadratic GUP, namely
%
%
\be\label{gulp}
\de x\,\de p \ \geq \ \frac{\hbar}{2}\lf(1 \ + \ \beta\,\frac{\de p^2}{m_p^2}\ri)\, ,
\ee
where $c=1$, $m_p =\sqrt{\hbar c/G} \approx 2.2\times 10^{-8}$kg is the Planck mass and $\beta$ is a dimensionless deformation parameter.
The symbol $\delta$ denotes uncertainty of a given observable
and it does not need to be {\em a priori} related to the standard deviation. More like in the original Heisenberg
uncertainty relation $\delta$ can represent Heisenberg's ``ungenauigkeiten'' (i.e., error-disturbance uncertainties caused by the backreaction in simultaneous measurement of $x$ and $p$) or $\delta p = \langle \psi ||p| | \psi \rangle \equiv \langle |p| \rangle_{\psi} $, see, e.g.~\cite{Maggiore:93}.

The quadratic GUP (\ref{gulp}) has served as an incubator for a number of important studies
in quantum mechanics~\cite{Bernardo,panella,ijmpd}, particle physics~\cite{Das,Husain},
finite-temperature quantum field theory~\cite{Scardigli_2} or cosmology~\cite{JS}.
In addition, the mass parameter in (\ref{gulp}) does not need to be necessarily
$m_p$  but it might be
identified with a characteristic mass scale in the effective quantum description, e.g.,
in condensed matter and atomic physics or in non-linear optics~\cite{Tawfik,Braidotti,JKS}.


In cases when $\delta$ represents the standard deviation (henceforth denoted as $\Delta$), the GUP inequality (\ref{gulp})
can be deduced from the deformed (Jacobi identity satisfying) commutation
relations (DCR)
\be\label{comm}
\lf[\hat{x},\hat{p}\ri] \ = \  i\hbar\lf(1 + \beta\,\frac{\hat{p}^2}{m_p^2}\ri)\,,\;\; \lf[\hat{x},\hat{x}\ri] \ = \ \lf[\hat{p},\hat{p}\ri] \ = \ 0\, ,
\ee
by means of the Cauchy--Schwarz or covariance inequality~\cite{Schrodinger,Robertson,Muk}, provided we focus  on
{\em mirror symmetric}
states where $\langle \hat{x}\rangle_{\psi}=\langle \hat{p}\rangle_{\psi}=0$, e.g., $\psi$ are parity eigenstates~\cite{adler}.

Most of recent discussions of GUP in quantum gravity have focused on heuristic applications in
cosmology and astrophysics (for review, see, e.g.~\cite{Tawfik}).
Comparably less attention has been devoted to study of GUP in a quasi-classical domain. It is, however, the quasi-classical
quantum regime that is pertinent in observational cosmology and astrophysics~\cite{Kiefer,Kiefer2}.
Important theoretical instruments used in quasi-classical quantum theory are coherent states (CS's).
This is because CS's are least susceptible to the loss of quantum coherence~\cite{Zurek:93}. In a sense,
CS's are the privileged states in the transition to classical reality, as they are the only states that
remain pure in the decoherence process~\cite{Dutra:97,Matacz:93}.

Various classes of CS's have been studied.
Here we will discuss the Schr\"{o}dinger-type minimum-uncertainty  CS's~\cite{Schrodinger,Nieto:78} associated with GUP.
%
%
%
%
%
%
We derive precise forms of such  GUP CS's both in the
momentum and position representation~\cite{SM:aa}, though our focus will be on the momentum representation
where CS's coincide with Tsallis probability amplitudes.
For $\beta  < 0$ we also reformulate the GUP in terms of one-parameter class of Tsallis
entropy-power based uncertainty relations (EPUR), which are saturated by the GUP CS's.
Since thermodynamics alongside with its various generalizations~\cite{jaynes1,jaynes2,Tsallis-book,kapur,abe,abe2} crucially hinges on the Maximum entropy principle (MEP) (i.e., thermodynamic entropy is the statistical entropy evaluated at the maximal entropy distribution),
we are led to the conclusion that the combination of GUP CS's with Tsallis entropy provides a natural framework to discuss the quasi-classical regime of GUP in terms of non-extensive thermostatistics of Tsallis (NTT)~\cite{Tsallis-book}.
We apply this insight to find ensuing GUP generalization of Verlinde's entropic gravity force and resulting cosmological implications in such a regime.

{\em Coherent states for GUP.}~---~We first summarize the steps leading to~(\ref{gulp}) from the DCR~(\ref{comm}).
To this end, we quantify the uncertainty of an observable $\mathcal{\hat{O}}$  with respect to a density matrix $\varrho$
via its standard deviation. In particular, for variance, i.e., square of the standard deviation we have
\begin{eqnarray}\label{var}
(\Delta\mathcal{\hat{O}})^2_{\varrho} &\equiv& \mbox{Tr}(\mathcal{\hat{O}}^2\varrho) -\mbox{Tr}(\mathcal{\hat{O}}\varrho)^2\ \nonumber \\[2mm]&=&
\int_{\mathbb{R}} \left(\lambda - \langle\mathcal{\hat{O}}\rangle^2_{\varrho} \right) \ \! d{\mbox{Tr}}\!\left(E_{\lambda}^{(\mathcal{\hat{O}})}\! \varrho \right).
\end{eqnarray}
Here $E_{\lambda}^{(\mathcal{\hat{O}})}$ is the projection–valued measure of $\mathcal{\hat{O}}$ corresponding to spectral value $\lambda$.
By confining our study to the observables $\hat{x}$ and $\hat{p}$, the passage from the DCR~(\ref{comm}) to GUP~(\ref{gulp}) is as follows:
we set $\mathcal{\hat{O}}_1 = \hat{x} - \langle \hat{x} \rangle_{\varrho} $ and $\mathcal{\hat{O}}_2 = \hat{p} - \langle \hat{p} \rangle_{\varrho}$ so that $(\Delta{{x}})^2_{\varrho} = \langle \mathcal{\hat{O}}_1^2\rangle_{\varrho} $, $(\Delta{{p}})^2_{\varrho} = \langle\mathcal{\hat{O}}_2^2\rangle_{\varrho} $ and $[\hat{x}, \hat{p}]\varrho = [\mathcal{\hat{O}}_1,\mathcal{\hat{O}}_2]\varrho$, then  for arbitrary vector $\psi \in {\mbox{Ran}} \varrho$ and any $\gamma \in \mathbb{R}$ we have
\begin{eqnarray}
0 &\leq& |\!|(\mathcal{\hat{O}}_2 -i\gamma \mathcal{\hat{O}}_1)\psi  |\!|^2 \nonumber \\[2mm]
&=& \langle \psi|  \mathcal{\hat{O}}_2^2 | \psi \rangle + i\gamma \langle \psi | [\mathcal{\hat{O}}_1,\mathcal{\hat{O}}_2]| \psi \rangle + \gamma^2 \langle \psi|  \mathcal{\hat{O}}_1^2 | \psi \rangle\, ,
\label{4.ab}
\end{eqnarray}
and therefore
\begin{eqnarray}
\mbox{Tr}(\mathcal{\hat{O}}_2^2 \varrho) + i\gamma \mbox{Tr}([\mathcal{\hat{O}}_1,\mathcal{\hat{O}}_2]\varrho)  + \gamma^2 \mbox{Tr}(\mathcal{\hat{O}}_1^2 \varrho) \ \geq \ 0\, .
\label{5.aac}
\end{eqnarray}
The LHS is smallest for $\gamma =  i \mbox{Tr}([\mathcal{\hat{O}}_2,\mathcal{\hat{O}}_1]\varrho) / (2 \mbox{Tr}(\mathcal{\hat{O}}_1^2 \varrho))$, which turns (\ref{5.aac}) to
\begin{eqnarray}
\mbox{\hspace{-3mm}}\mbox{Tr}(\mathcal{\hat{O}}_1^2 \varrho) \mbox{Tr}(\mathcal{\hat{O}}_2^2 \varrho)   =  (\Delta{{x}})^2_{\varrho} (\Delta{{p}})^2_{\varrho} \ \geq \ \frac{1}{4} \mbox{Tr}( i[\hat{x},\hat{p}]\varrho)^2.
\end{eqnarray}
This is nothing but quantum mechanical version of the {\em covariance inequality}
~\cite{Muk}.
Now we can use~(\ref{comm}) to obtain
\begin{eqnarray}
(\Delta{{x}})_{\varrho} (\Delta{{p}})_{\varrho} \ \geq \  \frac{\hbar}{2}\lf(1+\beta\,\frac{(\Delta{{p}})^2_{\varrho} + \langle \hat{p} \rangle^2_{\varrho}  }{m_p^2}\ri).
\label{split}
\end{eqnarray}
For {\em mirror symmetric} $\varrho$'s satisfying $\langle \hat{p} \rangle_{\varrho} = 0$
inequality~(\ref{split}) clearly coincides with the GUP~(\ref{gulp}), with variances in place of generic $\delta$'s.

To find  $\varrho$ that saturates the GUP (\ref{split}), we observe from~(\ref{4.ab}) that the inequality is saturated if and only if for all $\psi \in {\mbox{Ran}}\varrho$  the equation $(\mathcal{\hat{O}}_2 -i\gamma \mathcal{\hat{O}}_1)|\psi \rangle = 0$ holds. If this equation has for given $\gamma$, $\langle \hat{x} \rangle_{\varrho}$ and $\langle \hat{p} \rangle_{\varrho}$  more than one solution,
the corresponding minimum–uncertainty $\varrho$ is a mixture of CS's (i.e., pure minimum–uncertainty states).
It is apparent, cf. Eq.~(\ref{sol}) that on the class of mirror symmetric $\varrho$'s the equation
%
%
%
%
\be\label{eqn}
\lf(\hat{p}-i\gamma \hat{x}\ri)|\psi\rangle \ = \ 0\,,
\ee
has only one solution for $\psi \in L^2(\mathbb{R})$, so that the minimum–uncertainty $\varrho$  is a pure state --- CS.
It is convenient to seek the solution to (\ref{eqn}) in the momentum representation, i.e $|\psi\rangle \mapsto  \psi(p)=\langle p|\psi\rangle$.
In the momentum space, $\hat{x}$ and $\hat{p}$ satisfying DCR can be represented as in~\cite{kempf}.
%
%
However, by doing so, the non-symmetric nature of $\hat{x}$  would provide inconsistent variance for the ensuing CS, cf. Eq.~(\ref{var2}).
%
For this reason we resort to another representation of $\hat{x}$ and $\hat{p}$ complying with (\ref{comm}), namely
\begin{eqnarray}
&&\mbox{\hspace{-7mm}}\hat{p}\,\psi(p)\ = \ p\,\psi(p)\,, \nonumber \\[2mm]
&&\mbox{\hspace{-7mm}}\hat{x}\,\psi(p)\ = \ i\hbar\lf(\frac{d}{d p}+\frac{\beta}{2\,m_p^2}\,\lf[p^2,\frac{d}{d p}\ri]_+\ri)\psi(p)\,,
\label{operators}
\end{eqnarray}
with $[~,~]_+$ being an anticommutator.
With this, we  can cast~(\ref{eqn}) into an equivalent form
\be\label{diffeq}
\frac{d}{d p}\,\psi(p) \ = \ -\frac{\lf(1+\frac{\beta\ga\hbar}{m_p^2}\ri)}{\ga\hbar\lf(1  +  \beta\frac{p^2}{m_p^2}\ri)}\,p\,\psi(p)\,,
\ee
which admits the generic solution
\be
\psi(p) \ = \ N\left[1  +  ({\beta\,p^2})/{m_p^2}\right]^{-\frac{m_p^2}{2\beta\ga\hbar}-\frac{1}{2}}\, .
\label{sol}
\ee
The coefficient $N$ ensures that $\int|\psi(p)|^2dp=1$ and for $\beta >0$
%
\be\label{normal}
N_{_>} \ = \ \sqrt{\sqrt{\frac{\beta}{m^2_p\pi}}\,\frac{\Gamma\lf(\frac{m_p^2}{\beta\ga\hbar}+ 1\ri)}{\Gamma\lf(\frac{m_p^2}{\beta\ga\hbar} + \frac{1}{2}\ri)}}\,.
\ee
Here $\Gamma(x)$ is the gamma function ~\cite{grad}.

Situation with $\beta < 0$ has been less explored in literature than the $\beta > 0$ case,
though the related GUP has a number of important implications, e.g., in
cosmology~\cite{JKS,ours}, astrophysics~\cite{Ong} or DSR~\cite{MS1,MS2}.
Note that for $\beta < 0$  Eq.~(\ref{sol}) involves noninteger powers of negative reals, which lead to multi-valued CS. Because wave functions must be single-valued, CS has to have bounded support, which in turn means that $\hat{p}$ must be bounded with spectrum $|\sigma(\hat{p})|\leq m_p/\sqrt{|\beta|}$. The ensuing operator $\hat{x}$ corresponding to the formal differential expression (\ref{operators}) is self-adjoint~\cite{SM:aa}.
Resulting CS reads
\begin{eqnarray}
\psi(p) \ = \ N_{_<}\lf[1  - ({|\beta|\,p^2})/{m_p^2}\ri]_{+}^{\frac{m_p^2}{2|\beta|\ga\hbar}-\frac{1}{2}}\, ,
\label{sol2}
\end{eqnarray}
where $[z]_{+} = \max\{z,0 \}$ with
\begin{eqnarray}
N_{_<} \ = \ \sqrt{\sqrt{\frac{|\beta|}{m^2_p\pi}}\,\frac{\Gamma\lf(\frac{1}{2} + \frac{m_p^2}{|\beta|\ga\hbar}\ri)}{\Gamma\lf(\frac{m_p^2}{|\beta|\ga\hbar}\ri)}}.
\end{eqnarray}
In passing, we observe that as $\beta \rightarrow 0$  both~(\ref{sol}) and (\ref{sol2}) reduce to the usual minimum uncertainty Gaussian wave-packet (Glauber coherent state)
associated with the conventional Heisenberg uncertainty relation.

To find a physical meaning for $\gamma$, we note [see sentence after~(\ref{5.aac})] that for CS $\psi$
%
%
\begin{eqnarray}
\gamma &=&
-i \langle [\hat{x},\hat{p}] \rangle_{\psi}/[2(\Delta x)^2_{\psi}]
\ = \    - 2(\Delta p)^2_{\psi}/ i \langle [\hat{x},\hat{p}] \rangle_{\psi} \nonumber \\[2mm] &=& \  \frac{(\Delta p)_{\psi}}{(\Delta x)_{\psi}} \ = \ \frac{2(\Delta p)^2_{\psi}}{\hbar\lf[1 \ + \ \beta\,{(\Delta p)^2_{\psi}}/{m_p^2}\ri]}\,,
\label{14.abd}
\end{eqnarray}
where in the second and third  identity we utilized the fact that $\psi$ saturates~(\ref{split}).
%
Note also that CS's (\ref{sol})  satisfy  $\langle \hat{p} \rangle_{\psi} = \langle \hat{x} \rangle_{\psi} = 0$.

{\em Tsallis distribution.}~---~Let us now consider the following substitutions (valid for $\beta \lessgtr 0$) in (\ref{sol}) and (\ref{sol2}):
\begin{eqnarray}\label{qb}
q \ = \ \frac{\beta \gamma \hbar}{m^2_p+\beta\ga\hbar} + 1\,, \qquad  b \ = \ \frac{2 m_p}{\gamma \hbar}+\frac{2\beta}{m_p}\, .
\label{17.ccd}
\end{eqnarray}
%
With this, we can rewrite (\ref{sol}) and (\ref{sol2}) as
\begin{eqnarray}\label{finalsol}
\psi(p) \ = \ {N}_{_{\lessgtr}} \left[1 - b\,(1-q) \frac{p^2}{2 m_p} \right]^{\frac{1}{2(1-q)}}_+\, .
\label{CS-1}
\end{eqnarray}
This is nothing but the probability amplitude for the
Tsallis distribution of a free non-relativistic particle
\begin{eqnarray}
\mbox{\hspace{-2mm}}q_{_{T}}(p|q,b)  =   |\psi(p)|^2  =  \frac{1}{Z}\left[1  -  b\,(1-q) \frac{p^2}{2 m_p}\right]^{\frac{1}{1-q}}_+\!\! ,
\label{15aa}
\end{eqnarray}
with $Z= {N}_{_{\lessgtr}}^{-2}$ being the ``partition function''.

A few remarks concerning (\ref{15aa}) are now in order. Tsallis distribution of this type is also known as $q$-Gaussian distribution and denoted as $\exp_q(-b\,p^2/2\,m_p)$.
In the limit $q\rightarrow 1$, $\exp_q(-b\,p^2/2\,m_p) \rightarrow \exp(-b\,p^2/2\,m_p)$.  Note that because of~(\ref{qb}) $q\rightarrow 1$ is equivalent to $\beta \rightarrow 0$. In addition,  since for $\beta > 0$ the $\hat{p}$ operator is unbounded, CS~(\ref{finalsol}) is  normalizable only for values of $1\leq  q<3$. For values $q<1$ (i.e. $\beta < 0$), the distribution (\ref{15aa}) has a finite support with $|p|<\sqrt{2\,m_p/b\,(1-q)}$.
%
Moreover,  for $q \geq 5/3$ the variance of (\ref{15aa}) is undefined (infinite)~ and thus the GUP cannot even be formulated. When $q < 5/3$, then (see, e.g.~Ref.~\cite{Thistleton})
%
%
\begin{eqnarray}
\mbox{\hspace{-5mm}}&&(\Delta p)^2  =  \frac{2\,m_p}{b\,(5-3q)} \  \Leftrightarrow  \   \gamma  =  \frac{2(\Delta p)^2}{\hbar\lf[1  +  \beta{(\Delta p)^2}/{m_p^2}\ \!\ri]},
\label{var2}
\end{eqnarray}
which  coincides with (\ref{14.abd}) [this, in turn, justifies our choice of the representation of $\hat{x}$ and $\hat{p}$ operators].
Furthermore, the mean value does not exist for $q>2$, so such CS cannot be mirror-symmetric. Thus, the only physically relevant domain of $q$ in CS is $q < 5/3$, which ensures that $\beta$ is monotonically increasing function of $q$  and that $\beta >  -m^2_p/[3 (\Delta p)^2_{\psi}]$.


{\em Connection with entropic uncertainty relations.}~---~
Probability distribution (\ref{15aa}) decays
asymptotically following power law.
If variance and mean are the only
observables, power-law type distributions are incompatible with the conventional MEP
based on Shannon--Gibbs' entropy (SGE).
Nonetheless, distribution (\ref{15aa}) is a maximizer of Tsallis
(differential) entropy (TE) ${\mathcal{S}}_{2-q}^T$, where
%
\begin{eqnarray}
{\mathcal{S}}_q^T(\mathcal{F}) &=& \frac{k_B}{(1-q)}\left(  \int_{\mathbb{R}}d{p}\,\mathcal{F}^{q}(p)  - 1 \right)\, ,
\end{eqnarray}
($\mathcal{F}$ is a probability density function) subject to a constraint $\langle \hat{p}^2 \rangle_{\psi} = 2m_p/[b(5-3q)] $, cf.~\cite{Tsallis-book,Tsallis2,JKZ,JK}. $k_B$ is the Boltzmann constant.
In the limit $q \rightarrow 1$, the TE  tends to SGE
by L'H{o}pital's rule.

When dealing with GUP that is saturated by Tsallis CS, it is convenient to employ the concept of Tsallis entropy power (TEP)~\cite{JK:PRE}.
TEP $M_q^T$ of a random vector ${\mathcal{X}}$ is the
unique number that solves the equation
\begin{eqnarray}
{\mathcal{S}}_q^T
\left( {\mathcal{X}} \right)
\ = \ {\mathcal{S}}_q^T\left(\sqrt{M_q^T(\mathcal{X})}\cdot
{\mathcal{Z}}^{T}\right)\, .
\label{3.1.0kl}
\end{eqnarray}
Here ${\mathcal{Z}}^{T}$ represents a Tsallis random vector with zero mean
and unit covariance matrix. Such a vector is distributed with respect to the Tsallis distribution that extremizes ${\mathcal{S}}_q^T$. In Supplemental Material~\cite{SM:aa} we use Beckner--Babenko theorem~\cite{JK:PRE} to prove that for $\beta <0$ the DCR~(\ref{comm}) implies  the following one-parameter class of EPURs
\begin{eqnarray}
\mbox{\hspace{-5mm}}M_{q/2}^T(|{\psi}|^2) M_{1/(2- 2/q)}^T(|\tilde{\psi}|^2)   \ \geq \ \hbar^{2}/4\, ,  \;\;\; q \in [1,2)\, .
\label{PRL.29.ab}
\end{eqnarray}
Here $\tilde{\psi}$ is the position-space wave function associated with  ${\psi}$.
The clear advantage of EPUR (\ref{PRL.29.ab}) over GUP (\ref{gulp}) is in that the RHS has an irreducible and state-independent lower bound. Moreover,  (\ref{PRL.29.ab}) is also saturated by the GUP CS's~\cite{SM:aa}.

Numerical simulations based the Markovian master equations for reduced density matrix coupled with predictability sieve method~\cite{Zurek:93,Zurek:99,Venugopalan:99,Dalvit:05} indicate that CSs belong among the so-called {\em pointer states}, i.e. states that are least affected by the interaction with the environment (external degrees of freedom).
Such states belong to the quasi-classical domain of quantum theory as they are maximally predictable despite of decoherence~\cite{Dalvit:05,Zurek:01}.
Among all pointer states in the would-be GUP driven universe, CS's (\ref{CS-1}) have the highest TE. Moreover, EPUR~(\ref{PRL.29.ab}) indicates that TE is at the same time, a pertinent entropy functional in the GUP context.  So,  when we want to discuss a statistical physics of an ensemble of non-interacting GUP particles that are monitored by
quantum gravitational environment (bath of gravitons) we might invoke, similarly as in conventional statistical physics, MEP  but this time with TE in place of SGE. Ensuing NTT~\cite{Tsallis-book} can be then used to probe the quasi-classical
domain of GUP.
%
%
We now illustrate potential implications of this observation
%
with few examples.

{\em Examples.}~---~First, we consider modifications to Newton's law that should be expected in the quasi-classical epoch of GUP-based universe with $\beta\! < \!0$. To that end, we employ Verlinde's idea that gravity is an entropy-driven phenomenon ---  entropic gravity (EG)~\cite{verlinde}, alongside with
the NTT~\cite{Tsallis-book,abe}.


Following Verlinde's EG, we suppose that true (unknown) microscopic degrees of freedom in any given part of space are
stored in discrete bits on the holographic screen that surrounds them. A holographic screen can be considered to be spherically symmetric of area $A=4\pi R^2$. Outside of the screen is the {\em emergent world}, so the screen acts as an interface between known and unknown physics.
When a test particle moves away from the screen, it feels an effective force $F$ satisfying $F\delta x = T \delta S$, where $T$ and  $\delta S$ are temperature and the entropy change on the holographic surface, respectively and $\delta x$  is the distance of the particle from the screen. It should be stressed that Verlinde's thermodynamic relation is not directly related to the interior of the screen --- it operates in the emergent world. In NTT  the heat one-form $T \delta S$ must be replaced with~\cite{abe,abe2}: ${T\delta \mathcal{S}^T_q}/{[1+((1-q)/k_B)\mathcal{S}_{q}^T]}$ (in our context $\mathcal{S}_{q} \mapsto \mathcal{S}_{2-q}$). If $L$ is a (dimensionless) characteristic length scale (e.g. radius $R/\ell_p$)
then the Bekenstein--Hawking entropy $S_{BH} = \ln W(L) \propto L^2$, which implies that the total number of internal configurations $W$
behaves  for $L\gg 1$ as $W(L) = \phi(L) \nu^{L^2}$, where $\phi$ is any positive function satisfying $\lim_{L\rightarrow \infty} \ln\phi/L^2 =0$ and $\nu >1$ is some constant~\cite{tbh}. So, from the outside the holographic screen has entropy
\begin{eqnarray}
\mbox{\hspace{-5mm}}\mathcal{S}_{2-q}^T  \!= \!\  k_B\ln_{2-q} W(L)  \!=\!  \frac{k_B}{q-1}\left[\left(\phi(L) \nu^{L^2}\right)^{q-1}\!\! - 1 \right]\! .
\end{eqnarray}
Consequently, the entropic force follows from
\begin{eqnarray}
\mbox{\hspace{-5mm}}F\delta x
&=&  \frac{T\delta S_{2-q}^T}{1 + (q-1)\left(\omega_3 L^3 + \omega_2L^2  + \cdots\right) + \cdots }\, ,
\label{entfor}
\end{eqnarray}
where $\omega_2, \omega_3 >0$ are  intensive coefficients known from Hills' entropy expansion in (conventional) thermodynamics of small and mesoscopic systems~\cite{Hill}.
To comply with Hills' expansion we have formally included term $\omega_3 L^3$  even if it is not supported by  EG prescription. It will be seen that such a term is cosmologically unfeasible in the quasi-classical regime, so that $\omega_3 \approx 0$.
%

By holographic scaling, the energy residing inside the holographic screen is related with the on-screen degrees of freedom  via the equipartition theorem $E=Nk_BT/2$, with $E=M$ being the total mass enclosed by the surface and  $N=A/(G\hbar)$ the number of bits connected with the area by the holographic principle~\cite{verlinde}.

EG paradigm posits that the minimum possible
increase in the screen entropy (equivalent to one bit of Shannon's information) happens if a particle of radius of Compton wave length $\lambda_C$ is added to a holographic sphere~\cite{verlinde,Bekenstein}. This happens when a point-like quantum particle appears at the
distance $\lambda_C$ from the screen~\cite{note2}. By setting $\delta x=\lambda_{{C}}=\hbar/m$
and using the non-extensive version of Landauer principle~\cite{landauer,landauer-2},
which states that the erasure of information leads to an entropy increase $\delta S_q^T=2\pi k_B/(3-2q)$ per erased bit, we derive
the following modified Newton's law
%
%
\be\label{newton}
F(R)\ = \ \frac{GMm}{w R^2}\frac{1}{1-\kappa_3\varepsilon_q R^3 -\kappa_2\varepsilon_q R^2}\,,
\ee
with $\varepsilon_q = 1-q$, $w= 1 + 2\varepsilon_q$ and $\kappa_n = \omega_n/\ell_p^n$, $n=2,3$. Since $2\varepsilon_q$ is  small
(see below), we can set $w =1$.  Ensuing gravitational potential up to the first-order in $\varepsilon_q$ is
\begin{eqnarray}
&&\mbox{\hspace{-5mm}}V(R) \ = \ {\frac{r_s}{2}}\left[ -\frac{1}{R} \ + \ \varepsilon_q \kappa_2 R  \ + \  \frac{{\varepsilon_q\kappa_3}}{2} \ \!R^2  \right],
\label{newton2}
\end{eqnarray}
where $r_s = 2 GM $ is the Schwarzschild radius. Eq.~(\ref{newton2}) formally coincides with Mannheim--Kazanas {\em external} gravitational potential of a static, spherically symmetric source of mass $M$ in conformal gravity (CG)~\cite{mannheim,mannheim_a,mannheim_b,mannheim_c}.
%
Strictly speaking, in CG a given local gravitational
source generates only a gravitational potential
\begin{eqnarray}
V_{MK}(R)\ = \  -\frac{r_s }{2 R} \ + \ \frac{\chi }{2} R \, .
\label{newton3}
\end{eqnarray}
The would-be term $\propto R^2$ corresponds to a trivial vacuum solution of CG and hence does not couple to matter sources~\cite{mannheim,mannheim_a,mannheim_b,mannheim_c}.
Fitting with CG thus implies that $\omega_3 \approx 0$.
The magnitude of the constant $\chi$ can be associated with
the inverse Hubble radius~\cite{Kazanas:12}, i.e. $\chi \simeq 1/R_H$.
One should point out that by means of  $V_{MK}$ it has been successfully fitted more than two hundred galactic rotation curves (with no need for dark matter or other exotic modification of gravity)~\cite{mannheim,mannheim_a,mannheim_b,mannheim_c}.
Besides CG, the spherically-symmetric gravitational potential with a linear potential also occurs, e.g., in the dilaton-reduced action of gravity~\cite{5,6} or $f(R)$ gravity~\cite{11}.
%
%
%

To be more quantitative, let us assume that the GUP particle in question is inflaton. In such a case a quasi-classical (decoherence)
description  is valid at the late-inflation epoch
(after the first Hubble radius crossing)
and perhaps even after its end during reheating~\cite{KieferII,Burges}.
In this period the NTT should be a pertinent framework for the
description  of the ``inflaton gas''.  E.g., by viewing the ``inflaton gas'' as the ideal gas
NTT predicts that the inflaton pressure should satisfy for $0< q < 1$ a polytrope relation $p \propto \rho^{5/3}$ ($\rho$ is energy density)~\cite{Abe4,Du:04}. Relation of this type frequently appears in phenomenological studies on late inflation~\cite{Barrow:16,Barrow:18}.
We can fix $\beta$ by matching
the linear terms in Eqs.~\eqref{newton2} and~\eqref{newton3}.
By using $r_s \simeq  R_H$, $\kappa_2 = \pi/ \ell_p^2\ \!$cm$^{-2}$ (Bekenstein--Hawking value) we get $\varepsilon_q = \ell_p^2/ (\pi R_H^2)$. Note that in this
setting the effect of the linear potential is comparable to that
of the Newtonian potential on length scales $1/R^2 \simeq \varepsilon_q \kappa_2$, i.e. $R \simeq \ell_p \sqrt{1/(\pi \varepsilon_q)} = R_H$.
By solving~(\ref{17.ccd}) with respect to $\beta$ and employing~\eqref{14.abd}, we obtain $|\beta| \simeq m_p^2\, \ell_p^2/(2\pi\left(\Delta p\right)^2_\psi R_H^2)$.
To estimate  $\beta$,
we express the Hubble radius as $R_H(t)=H^{-1}(t)=a(t)/\dot{a}(t)$,
where $H$ is the Hubble parameter
and the scale factor $a(t)$ can be evaluated from the Vilenkin--Ford model~\cite{VFmodel}:
$a(t)=A\sqrt{\mathrm{sinh}\left(B\,t\right)}$, with $B=2\sqrt{\Lambda/3}$
($\Lambda$ is the cosmological constant).
On the other hand, from the relativistic equipartition theorem we have
$\left(\Delta p\right)^2_\psi \simeq 12 \left(k_B T\right)^2$, cf.~\cite{SM:aa}.
A straightforward computation
gives
\be
\label{betaest2}
|\beta| \ \equiv \ |\beta|(t)\ = \ \frac{m_p^2\, \ell_p^2\,\Lambda}{72\pi\left(k_B T\right)^2\mathrm{tanh}^2\left({2t\sqrt{\frac{\Lambda}{3}}}\right)}\,.
\ee
%
%
For concreteness' sake, let us consider the late-inflation/reheating epoch,
i.e. time scale $t\simeq 10^{-33}$s. By assuming
$T$ of the order of the reheating temperature $T_R\simeq 10^{7} \div 10^8$GeV, we obtain $|\beta| \sim  10^{-2} \div 1$, which is in agreement with the values predicted by string theory, cf. e.g.~\cite{string,string II}.
In passing we stress that the above connection with the CG potential works only for $\beta<0$, or else in~\eqref{newton2} we would have a wrong sign in front of the linear potential.

We finally note that the DCR (\ref{comm}) with $\beta <0$ given by (\ref{betaest2})
is consistent with Magueijo--Smolin DSR~\cite{MS1,MS2}.
In a nutshell, DSR is a theory which coherently tries to implement a second invariant (namely $m_p$ or equivalently $\ell_p$),
besides the speed of light, into the transformations among inertial reference frames. The Magueijo--Smolin DSR model predicts
that the DCR should vanish at Planck scale (thus physics should be deterministic there) while at low energies it approaches the
conventional canonical commutator. In our case
we indeed see from (\ref{betaest2}) that for allowed cosmological times $|\beta|(t)$ monotonically decreases with increasing $t$.


{\em {Conclusions.}}~---~To conclude, we have derived explicit form of coherent states for generalized uncertainty principle and showed that in the momentum representation they coincide with Tsallis probability amplitudes. Furthermore, for $\beta < 0$, we have reformulated GUP in terms of Tsallis-entropy based entropic uncertainty relations and by invoking the Maximal Entropy principle, we showed that in the semi-classical (decoherence) limit one can establish equivalence between the GUP quantum systems and non-extensive thermostatistics of Tsallis. This
provides a novel framework to discuss transition between the GUP quantum substrate and classical reality and opens a viable route for table-top experiments to explore possible GUP-based quantum gravitational phenomena via analog gravity models.

%

\begin{acknowledgments}
{\em {Acknowledgments.}}~---~We thank Vladimir Lotoreichik
and Pavel Exner for useful discussions and comments on the manuscript. P.J. was supported by the Czech Science Foundation Grant (GA\v{C}R), Grant 19-16066S.
\end{acknowledgments}


\begin{thebibliography}{99}
\bibitem{note2} Note that $\lambda_C$ is the  minimal distance at which particle still retains its single particle picture.
\end{thebibliography}
\end{document}



\title{Supplemental Material for ``Decoherence limit of quantum systems obeying generalized uncertainty principle:
new paradigm for Tsallis thermostatistics''}
%

\author{Petr Jizba}
\email{p.jizba@fjfi.cvut.cz}
\affiliation{FNSPE,
Czech Technical University in Prague, B\v{r}ehov\'{a} 7, 115 19, Prague, Czech Republic}
\affiliation{ITP, Freie Universit\"{a}t Berlin, Arnimallee 14, D-14195 Berlin, Germany}

\author{Gaetano Lambiase}
\email{lambiase@sa.infn.it}
\affiliation{Dipartimento di Fisica, Universit\`a di Salerno, Via Giovanni Paolo II, 132 I-84084 Fisciano (SA), Italy}
\affiliation{INFN, Sezione di Napoli, Gruppo collegato di Salerno, Italy}

\author{Giuseppe Gaetano Luciano}
\email{gluciano@sa.infn.it}
\affiliation{Dipartimento di Fisica, Universit\`a di Salerno, Via Giovanni Paolo II, 132 I-84084 Fisciano (SA), Italy}
\affiliation{INFN, Sezione di Napoli, Gruppo collegato di Salerno, Italy}

\author{Luciano Petruzziello}
\email{lupetruzziello@unisa.it}
\affiliation{Dipartimento di Fisica, Universit\`a di Salerno, Via Giovanni Paolo II, 132 I-84084 Fisciano (SA), Italy}
\affiliation{INFN, Sezione di Napoli, Gruppo collegato di Salerno, Italy}
\affiliation{Dipartimento di Ingegneria, Universit\`a di Salerno, Via Giovanni Paolo II, 132 I-84084 Fisciano (SA), Italy}


\vspace{2mm}
\maketitle

{\bf Note:} equations and citations that are related to the main text are shown in red.

%
%

\section*{A. Eigenstates of the position operator}
\label{sa}

Because the canonical commutator has the form {\cdred(2) }, the plane waves are not any more eigenstates of the operator $\hat{x}$ in the momentum representation and $\hat{p}$ in the position representation. This in turn means  that weave functions in position and momentum representation are not connected via Fourier transform. To see how they are related, let us first consider the eigenstates
of the operator $\hat{x}$ in the momentum representation. This is given by solving the following eigenvalue  equation
\begin{eqnarray}
\hat{x}\langle p|x\rangle \ \equiv \ x \psi_x(p) \ &=& \ i\hbar \left(\frac{d}{dp} + \frac{\beta}{m^2_p} p^2 \frac{d}{dp} + \frac{\beta}{m^2_p} p\right)\psi_x(p)\, ,
\end{eqnarray}
that can be equivalently rewritten as
\begin{eqnarray}
\frac{d}{dp} \ \! \psi_x(p) \ = \ \frac{\left(x \ - \ i\hbar \frac{\beta}{m^2_p} p\right)}{i\hbar\left(1 \ + \ \frac{\beta}{m^2_p} p^2 \right)} \ \! \psi_x(p) \, .
\end{eqnarray}
This has the solution
\begin{eqnarray}
\psi_x(p) \ = \ A_x \ \! \frac{e^{-ixm_p\arctan\left(p \sqrt{\beta}/m_p  \right)/\hbar\sqrt{\beta}}}{\sqrt{m^2_p  \ + \ p^2 \beta}}\, ,
\label{3aa}
\end{eqnarray}
for positive $\beta$, and
\begin{eqnarray}
\psi_x(p) \ = \ B_x \ \!
\frac{e^{-ixm_p{\mbox{\scriptsize{arctanh}}}\left(p \sqrt{|\beta|}/m_p  \right)/\hbar\sqrt{|\beta|}}}{\sqrt{m^2_p  \ - \ p^2 |\beta|}}\, ,
\label{3ab}
\end{eqnarray}
for negative $\beta$ (i.e. $\beta = -|\beta|$)

Let us now discuss the two cases ($\beta >0$ and $\beta <0$) separately.

\subsection{
Positive $\beta$ case}

In this case $\psi_x(p)$  is quadratically integrable and  the normalization factor $A_x$ can be chosen so as to ensure the normalizability to 1. It is easy to see that
\begin{eqnarray}
A_x \ = \ \sqrt{\frac{m_p}{\pi}} \beta^{1/4}\, .
\label{4aa}
\end{eqnarray}
It should be noted that in the limit $\beta \rightarrow 0$ the eigenstate (\ref{3aa}) with $A$ given by (\ref{4aa}) converges to zero and not to $e^{-ixp/\hbar}/\sqrt{2\pi \hbar}$ due to different normalization (plane waves are not quadratically integrable). On the other hand when $A_x$ is not considered then $\lim_{\beta\rightarrow 0} \psi_x(p)$ is proportional to a plane wave, namely $e^{-ixp/\hbar}/m_p$.

Should $\hat{x}$ be self-adjoint, the corresponding eigenfunctions $\psi_x(p)$ must be orthogonal for different $x$.
By requiring that $x$ is real and the ensuing  eigenfunctions $\psi_x(p)$ are orthogonal,  the completeness relation reads
\begin{eqnarray}
\langle x| x' \rangle \ = \ \int_{\mathbb{R}} dp \ \! \psi_{x}^*(p) \psi_{x'}(p) &=&  \int_{\mathbb{R}} dp \ \! \frac{\sqrt{\beta }m_p}{\pi}\frac{e^{-i(x'-x)m_p\arctan\left(p \sqrt{\beta}/m_p  \right)/\hbar\sqrt{\beta}}}{{m^2_p  \ + \ p^2 \beta}}\nonumber \\[2mm]
&=& \int_{-\frac{m_p\pi}{2\hbar \sqrt{\beta}}}^{\frac{m_p\pi}{2\hbar \sqrt{\beta}}} dz \ \! \frac{\sqrt{\beta }\hbar}{m_p\pi} \ \! e^{-i (x'-x) z} \ = \ \frac{2\hbar \sqrt{\beta }}{m_p\pi (x'-x)} \ \! \sin\left(\frac{m_p\pi (x'-x)}{2\hbar \sqrt{\beta }} \right)\, ,
\end{eqnarray}
where $\psi_{x}^*(p) = \langle x| p\rangle \equiv \psi_p(x)$. So, $x$ can acquire only discrete values $n 2\hbar \sqrt{\beta }/m_p + x_0$ with $n \in \mathbb{Z}$ and $x_0 \in (0, 2\hbar \sqrt{\beta }/m_p)$. In other words, we have a one-parameter family of eigenvectors
$\{\psi_{x \ \! = \ \! n (2\hbar \sqrt{\beta }/m_p) + x_0}(p)\}_{n \in \mathbb{Z}}$ parametrized by $x_0$.
Note that the difference between nearest eigenvalues is $2\hbar \sqrt{\beta }/m_p = 2(\delta x)_{\rm min} = 2(\Delta x)_{\rm min} $, which  goes to zero as $\beta \rightarrow 0$. Note that this $\delta x_{\rm min}$  coincides with the minimal positional variance implied by {\cdred(1)}.
In contrast to $(\Delta x)_{\rm min} $, the GUP {\cdred(1)} does not provide any universal bound for $(\Delta p)_{\rm min}$ or $(\Delta p)_{\rm max}$.

It should be noticed that $\hat{x}$ is not self-adjoint on the natural domain
\begin{eqnarray}
\mathcal{D}_x({\mathbb{R}})\ = \ \left\{\psi \in L^2({\mathbb{R}}): \psi \in  {AC}(\mathcal{\mathbb{R}}), \ \! \hat{x}\psi \in   L^2({\mathbb{R}})  \right\}\, ,
\end{eqnarray}
($AC({\mathbb{R}})$ stands for {\em absolutely continuous functions} on $\mathbb{R}$).
This is because $\hat{x}$ has eigenvalues $\pm i$
with eigenstates belonging to $\mathcal{D}_x({\mathbb{R}})$. Since the dimensionality of the respective Hilbert subspace is $1$, the {\em deficiency index} of such operator $\hat{x}$ is  $(1,1)$. It is possible to appropriately restricting the definition region of $\hat{x}$ so that the  deficiency index is $(0,0)$. For instance, we can require that $\psi(p=\infty) = \theta \psi(p=-\infty)$ where $|\theta| = 1$.  So, if we define $\mathcal{D}_x^{\theta}({\mathbb{R}})\ = \ \{\psi \in \mathcal{D}_x({\mathbb{R}}): \psi(\infty) = \theta \psi(-\infty), |\theta| =1\}$ we obtain a bijective correspondence between symmetric extension of the operator $\hat{x}$ on $L^2({\mathbb{R}})$ and complex numbers with $|\theta| = 1$; each of these operators $\hat{x}_{\theta} = \hat{x} \upharpoonright \mathcal{D}_x^{\theta}({\mathbb{R}})$ is self-adjoint.
with a discrete spectrum. In addition, by writing $\theta = e^{i\alpha}$ a setting $\alpha =  x_0m_p \pi/(\hbar \sqrt{\beta})$ the spectrum of $\hat{x}_{\theta} = n 2\hbar \sqrt{\beta }/m_p + x_0$, with $n \in \mathbb{Z}$,.

It should be stressed that the discreteness of the spectrum for the operator $\hat{x}$ is not compatible with $\beta >0$. Indeed, the expectation value of the canonical commutation relation {\cdred (2)} with respect to any eigenstate of $\hat{x}$ gives a zero left-hand side of {\cdred (2)}, while the right-hand side is always non-zero for $\beta>0$. Such a situation would not happen should both $\hat{p}$ and $\hat{x}$ have a continuous spectrum because then the corresponding eigenstates do not belong to the
domain of the commutator (see next subsection).
%
 %
On the other hand, the expectation value of the right-hand side of {\cdred (2)} can be zero but only if $\beta < 0$ but this would contradict our original assumption that $\beta >0$.

\subsection{
Negative $\beta$ case}

In this case $\psi_x(p)$  is not quadratically integrable. Indeed, the corresponding integral
\begin{eqnarray}
|\!| \psi_x|\!|^2 \ &=& \ \langle x|  x\rangle \ =  \ |B_x|^2 \int_{-m_p/\sqrt{|\beta|}}^{m_p/\sqrt{|\beta|}} \frac{dp }{m_p^2 - p^2|\beta|} \ = \ \{ z =  m_p\ \! \mbox{{arctanh}}\left(p \sqrt{|\beta|}/m_p  \right)/\sqrt{|\beta|}\}\nonumber \\[2mm]
&=& \ \left(\frac{|B_x|}{m_p}\right)^2 \int_{-\infty}^{\infty} dz \ = \ \infty\, .
\end{eqnarray}
Ensuing scalar product for two eigenstates is
\begin{eqnarray}
\langle x'|  x\rangle  &=&  {|B_x|}^2 \int_{-m_p/\sqrt{|\beta|}}^{m_p/\sqrt{|\beta|}}dp \ \! \frac{ e^{-i(x-x')m_p{\mbox{\scriptsize{arctanh}}}\left(p \sqrt{|\beta|}/m_p  \right)/\hbar\sqrt{|\beta|}}}{m_p^2 - p^2|\beta|} \ = \
\ \left(\frac{|B_x|}{m_p}\right)^2 \int_{-\infty}^{\infty} dz \ e^{-i(x-x')z/\hbar}\nonumber \\[2mm] &=& \left(\frac{|B_x|}{m_p}\right)^2 2\pi \hbar \ \!\delta(x-x')\, .
\end{eqnarray}
So, we can set $B_x = \sqrt{m_p^2/2\pi \hbar}$. Such eigenstates do not belong to the usual $L^2((-m_p/\sqrt{|\beta|},m_p/\sqrt{|\beta|}))$ space. Instead they belong to the space ${{\mathcal{S}}}'((-m_p/\sqrt{|\beta|},m_p/\sqrt{|\beta|}))$ of complex valued tempered distributions.
Hence, there is a spectral transition from discrete to continuous spectrum when $\beta$ becomes negative.
In the following subsection we show that the $\hat{x}$ operator is self-adjoint.

In passing we should note that although continuous observables such
as position $x$ or $p$ are routinely employed in quantum theory, they are
really unphysical idealizations: the set of possible
outcomes in any realistic measurement is always
countable, since the state space of any apparatus
with finite spatial extent has a countable
basis.
%
%
So, our reasonings related
to $\beta < 0$ should be thus understood in this mathematically idealized sense --- as done with conventional Heisenberg $p$-$x$ uncertainty relations.

The case with $\beta < 0$ is interesting from yet another point. While $\beta > 0$ predicts existence of $(\Delta x)_{\rm{min}} =  \sqrt{\beta} \hbar/ m_p =  \sqrt{\beta} \ell_p$ ($\ell_p = \sqrt{\hbar G/c^3}\approx 10^{-33}$cm is the Planck length) but does not provide any universal bound for $(\Delta p)_{\rm{min}}$ or $(\Delta p)_{\rm{max}}$ the case with  $\beta < 0$ allows only for momenta from the interval $(-m_p/\sqrt{|\beta|}, m_p/\sqrt{|\beta|})$
which implies that there exists $(\Delta p)_{\rm{max}} = m_p/\sqrt{|\beta|}$.

\subsection{Self-adjointness of the position operator ---  negative $\beta$ case}

First we note that the operator $\hat{p}$  defined in~{\cdred{(9)}} is the conventional {\em operator of multiplication} (by $p$), which is defined on the domain
\begin{eqnarray}
D(\hat{p}) \ = \ \left\{\psi \in L^2(\mathcal{I}):  p \ \! \psi \in   L^2(\mathcal{I})  \right\}\, ,
\end{eqnarray}
with $\mathcal{I}$ being an open interval with the endpoints $-a,a$  $(a = m_p/\sqrt{|\beta|})$.  Operators of multiplication are known to be dense (i.e., closure $\overline{D(\hat{p})} = L^2(\mathcal{I})$) and self-adjoint~\cite{Exner}.

Let us now show that the operator $\hat{x}$ corresponding to the {\em formal differential expression} {\cdred(9)} is self-adjoint.
The reason why we are interested in a self-adjointness of the position operator is that only self-adjoint operators (and not symmetric) have guaranteed real spectrum~\cite{footnote1}. To this end we define the  subspace
\begin{eqnarray}
\mathcal{D}_x(\mathcal{I})\ = \ \left\{\psi \in L^2(\mathcal{I}): \psi \in  {AC}(\mathcal{I}), \ \! \hat{x}\psi \in   L^2(\mathcal{I})  \right\}\, ,
\end{eqnarray}
($AC(\mathcal{I})$ stands for {\em absolutely continuous functions} on $\mathcal{I}$, see~\cite{footnote2}) which is  dense in $L^2(\mathcal{I})$ as it contains, e.g., set of  $C^{\infty}(\mathcal{I})$ functions that is dense in $L^2(\mathcal{I})$. $\mathcal{D}_x(\mathcal{I})$ is chosen so that it is the largest reasonable domain on which the operator $\hat{x}$ could be expected to act.

Let us now denote with a symbol $\hat{\mathsf{x}}$ the operator~\cite{footnote3}
\begin{eqnarray}
\hat{\mathsf{x}} \ \! \psi \ \equiv \ i\hbar \left(\frac{d}{dp} + \frac{\beta}{m^2_p} p^2 \frac{d}{dp} + \frac{\beta}{m^2_p} p\right)\psi\, , \;\;\; D(\hat{\mathsf{x}}) \ \equiv \ \mathcal{D}_x(\mathcal{I})\, .
\end{eqnarray}
We claim that this operator is self-adjoint when $\beta <0$.

Suppose first that $\psi$ and $\phi$ are arbitrary representatives from $D(\hat{\mathsf{x}})$. Then integration by parts gives
\begin{eqnarray}
\langle \psi| \hat{\mathsf{x}} \ \! \phi \rangle &=& i\hbar \int_{-a}^a dp \ \! \left[\psi^*(p){\phi'(p)}+ \frac{\beta}{m^2_p}p^2 \psi^{*}(p){\phi'(p)}  + \frac{\beta}{m^2_p} p \ \! \psi^{*}(p)\phi(p)\right]\nonumber \\[2mm]
&=& i\hbar |\![ \psi, \phi ]\!|_{a}  \ - \ i\hbar \int_{-a}^a dp \ \! \left[{\psi^{*}}'(p){\phi(p)}+ \frac{\beta}{m^2_p}p^2 {\psi^{*}}'(p){\phi(p)}  + \frac{\beta}{m^2_p} p \ \! \psi^{*}(p)\phi(p)\right]\nonumber \\[2mm]
&=& i\hbar |\![ \psi, \phi ]\!|_{a} \ + \ \langle \hat{\mathsf{x}} \ \! \psi| \phi \rangle\, ,
\label{12.cd}
\end{eqnarray}
where
\begin{eqnarray}
|\![ \psi, \phi ]\!|_{a} \ = \ \lim_{p \rightarrow \ \! a_{-}} \left(1 - \frac{|\beta|}{m^2_p} p^2\right)\psi^*(p)\phi(p) \ - \ \lim_{p \rightarrow \ \! -a_{+}} \left(1 - \frac{|\beta|}{m^2_p} p^2\right)\psi^*(p)\phi(p)\, .
\label{13.aa}
\end{eqnarray}
Note that both limits exist for any $\psi, \phi \in D(\hat{\mathsf{x}})$, even in the cases when the endpoints are singular or the one–sided limits of the functions $\psi, \phi$ in them make no sense. This is because for any $\psi, \phi \in D(\hat{\mathsf{x}})$  we have that $\psi^*(\hat{\mathsf{x}} \ \! \phi) - (\hat{\mathsf{x}} \ \! \psi)^*\phi$ belongs to $L^1(\mathcal{I})$, and so for any $d < a$
\begin{eqnarray}
\lim_{c\rightarrow -a_+} \int_{c}^{d} dp \ \! \left[ \psi^*(\hat{\mathsf{x}} \ \! \phi) - (\hat{\mathsf{x}} \ \! \psi)^*\phi \right] \ = \ \int_{-a}^{d} dp \ \! \left[ \psi^*(\hat{\mathsf{x}} \ \! \phi) - (\hat{\mathsf{x}} \ \! \psi)^*\phi \right] \ = \ \mbox{finite}\, .
\end{eqnarray}
Thus,
\begin{eqnarray}
\lim_{p \rightarrow \ \! -a_{+}} \left(1 - \frac{|\beta|}{m^2_p} p^2\right)\psi^*(p)\phi(p)\, ,
\end{eqnarray}
has in $-a$ a proper limit. Similarly for the $p\rightarrow a_-$ limit. So, both limits approach a finite constant, which by the convergence criterion for (improper) integrals must be zero~\cite{footnote4a}. This shows that $\hat{\mathsf{x}}$ is symmetric on the domain $D(\hat{\mathsf{x}})$.
From (\ref{12.cd}) we can thus conclude~\cite{footnote4} that every element $\varphi$ from $D(\hat{\mathsf{x}})$ is also in $D(\hat{\mathsf{x}}^{\dagger})$ (so that $D(\hat{\mathsf{x}}) \subset D(\hat{\mathsf{x}}^{\dagger})$), and then $\hat{\mathsf{x}}\varphi = \hat{\mathsf{x}}^{\dagger}\varphi$. That is, the adjoint of $\hat{\mathsf{x}}$ is an extension of $\hat{\mathsf{x}}$. To prove that $\hat{\mathsf{x}}$ is self-adjoint, we must show that every element in $D(\hat{\mathsf{x}}^{\dagger})$ is also in $D(\hat{\mathsf{x}})$. Thus, let $\varphi ,\eta \in  L^2(\mathcal{I})$, with $\langle \varphi| \hat{\mathsf{x}}\ \! \psi \rangle = \langle \eta| \psi \rangle $ for every $\psi \in D(\hat{\mathsf{x}})$. We must show that any such $\varphi$ is in ${AC}(\mathcal{I})$, with $\eta = \hat{\mathsf{x}}\varphi$. To this end we consider an arbitrary compact interval $\mathcal{I}' = [\mbox{$\mathfrak{a}$},\mbox{$\mathfrak{b}$}] \subset \mathcal{I}$ and
define on it
\begin{eqnarray}
\tilde{\eta}_{{\mathcal{I}}'}(p) \ = \ \frac{c}{\sqrt{1 + \frac{\beta}{m_p^2}p^2}} \ + \ \frac{1}{\sqrt{1 + \frac{\beta}{m_p^2}p^2}} \ \!  \int_{\footnotesize{\mbox{$\mathfrak{a}$}}}^{p} dt \ \! \frac{\eta(t)}{\sqrt{1 + \frac{\beta}{m_p^2}t^2} }\, ,
\label{SM.16.aa}
\end{eqnarray}
where the constant $c$ will be determined shortly. From (\ref{SM.16.aa}) it follows that  $\tilde{\eta}_{{\mathcal{I}}'}(p) \ \!\sqrt{1 + \frac{\beta}{m_p^2}p^2} \in {AC}(\mathcal{I}')$. Since also  $\sqrt{1 + \frac{\beta}{m_p^2}p^2}$ belongs to ${AC}(\mathcal{I}')$ and  $\sqrt{1 + \frac{\beta}{m_p^2}p^2} \neq 0$ on $\mathcal{I}'$, we have that $\tilde{\eta}_{{\mathcal{I}}'} \in {AC}(\mathcal{I}')$ and
\begin{eqnarray}
\left(\frac{d}{dp} + \frac{\beta}{m^2_p} p^2 \frac{d}{dp} + \frac{\beta}{m^2_p} p\right)\tilde{\eta}_{{\mathcal{I}}'}(p) \ = \ \eta(p)\, ,
\label{SM.17.ad}
\end{eqnarray}
holds in ${\mathcal{I}}'$ in the $L^2$ sense. Let now $\psi_{{\mathcal{I}}'}$  be an arbitrary function from
$D(\hat{\mathsf{x}})$ whose support is in ${\mathcal{I}}'$, then
\begin{eqnarray}
\langle \varphi| \hat{\mathsf{x}}\ \!  \psi_{{\mathcal{I}}'} \rangle  \ = \ \langle \eta| \psi_{{\mathcal{I}}'} \rangle &=& \int_{{{\mathcal{I}}'}} dp \ \!\left[\left(\frac{d}{dp} + \frac{\beta}{m^2_p} p^2 \frac{d}{dp} + \frac{\beta}{m^2_p} p\right)\tilde{\eta}_{{\mathcal{I}}'}  \right]^*\psi_{{\mathcal{I}}'} \nonumber \\[2mm]
&=& - \int_{{{\mathcal{I}}'}} dp \  \tilde{\eta}_{{\mathcal{I}}'}^* \left(\frac{d}{dp} + \frac{\beta}{m^2_p} p^2 \frac{d}{dp} + \frac{\beta}{m^2_p} p\right) \psi_{{\mathcal{I}}'} \ = \ \langle \tilde{\eta}_{{\mathcal{I}}'}| (i/\hbar) \hat{\mathsf{x}}\ \!  \psi_{{\mathcal{I}}'} \rangle  \, ,
\end{eqnarray}
which can be written as
\begin{eqnarray}
\langle \varphi_{{\mathcal{I}}'} + (i/\hbar) \ \!\tilde{\eta}_{{\mathcal{I}}'}| \hat{\mathsf{x}}\ \!  \psi_{{\mathcal{I}}'} \rangle  \ = \ 0\, ,
\label{SM.19.ac}
\end{eqnarray}
where $\varphi_{{\mathcal{I}}'}$ is the corresponding restriction of $\varphi$ from ${\mathcal{I}}$ to ${\mathcal{I}}'$ (i.e. in terms of characteristic function $\chi_{{\mathcal{I}}'}$ of a set ${\mathcal{I}}'$ we have $\varphi_{{\mathcal{I}}'} = \chi_{{\mathcal{I}}'} \varphi$).   At this stage we define the function $\zeta_{{\mathcal{I}}'}$  on ${\mathcal{I}}$, that is zero outside of   ${{\mathcal{I}}'}$ and for $p \in {{\mathcal{I}}'}$ it is defined by the expression
\begin{eqnarray}
\zeta_{{\mathcal{I}}'}(p) \ = \  \frac{1}{\sqrt{1 + \frac{\beta}{m_p^2}p^2}} \ \!  \int_{\footnotesize{\mbox{$\mathfrak{a}$}}}^{p} dt \ \! \frac{\left[\varphi_{{\mathcal{I}}'}(t) + (i/\hbar) \ \!\tilde{\eta}_{{\mathcal{I}}'}(t)\right]}{\sqrt{1 + \frac{\beta}{m_p^2}t^2} }\, .
\end{eqnarray}
Now we chose $c$ from (\ref{SM.16.aa}) so that $\zeta_{{\mathcal{I}}'}(\mbox{$\mathfrak{b}$}) = 0$. This ensures that $\zeta_{{\mathcal{I}}'} \in {AC}(\mathcal{I}')$. Besides
\begin{eqnarray}
\left(\frac{d}{dp} + \frac{\beta}{m^2_p} p^2 \frac{d}{dp} + \frac{\beta}{m^2_p} p\right) \zeta_{{\mathcal{I}}'}(p) \ = \ \varphi_{{\mathcal{I}}'}(p) + (i/\hbar) \ \!\tilde{\eta}_{{\mathcal{I}}'}(p)\, .
\label{SM.21.aa}
\end{eqnarray}
Since $\varphi_{{\mathcal{I}}'}$,  $\tilde{\eta}_{{\mathcal{I}}'}$ and $\zeta_{{\mathcal{I}}'}$ are absolutely continuous on a compact interval ${\mathcal{I}}'$ they belong to $L^2({\mathcal{I}})$ and due to (\ref{SM.21.aa}) $\zeta_{{\mathcal{I}}'}$ belongs to a set of function from $D(\hat{\mathsf{x}})$ with support in ${\mathcal{I}}'$. Because (\ref{SM.19.ac}) is valid for any  $\psi_{{\mathcal{I}}'} \in D(\hat{\mathsf{x}})$ with support in ${\mathcal{I}}'$ we can chose $\psi_{{\mathcal{I}}'} = \zeta_{{\mathcal{I}}'}$ in which case (\ref{SM.19.ac}) implies
\begin{eqnarray}
\varphi_{{\mathcal{I}}'} + (i/\hbar) \ \!\tilde{\eta}_{{\mathcal{I}}'} \ = \ 0\, .
\label{SM.22.cc}
\end{eqnarray}
Hence $\varphi_{{\mathcal{I}}'}  \in AC({\mathcal{I}}')$ and  $\hat{\mathsf{x}}\ \!\varphi_{{\mathcal{I}}'}(p) = \eta_{{\mathcal{I}}'}(p)$ by (\ref{SM.17.ad}) for almost all $p\in {\mathcal{I}}'$. Since $ {\mathcal{I}}'$ is an arbitrary compact interval
in $ {\mathcal{I}}$ it follows that $\hat{\mathsf{x}}\ \!\varphi = \eta$ almost everywhere in ${\mathcal{I}}$ and because  $\eta \in L^2({\mathcal{I}})$ we finally arrive at the desired result that $\varphi \in AC({\mathcal{I}})$ [by (\ref{SM.22.cc})] and  $\hat{\mathsf{x}}\ \!\varphi \in L^2({\mathcal{I}})$.

As an independent check we  compute the {\em deficiency indices} of the symmetric operator $\hat{\mathsf{x}}$. For symmetric operator $\hat{A}$ there are two deficiency indices defined as $n_{\pm}(\hat{A}) = \mbox{dim}\ \!{\mbox{Ker}}(\hat{A}^\dag \pm i)$ and usually written as the ordered pair $(n_+(A), n_-(A))$. A symmetric operator
$\hat{A}$ is self–adjoint if and only if it is closed and  $(n_+(A), n_-(A)) = (0,0)$, see, e.g.~\cite{Exner}.
Since $D(\hat{\mathsf{x}}^\dagger) = AC({\mathcal{I}})$   and $\hat{\mathsf{x}}^\dagger \varphi = i\hbar [\varphi'+ (\beta/m^2_p) p^2 \varphi' +  (\beta/m^2_p) p\varphi]$, the problem reduces to finding
solutions of the equation
\begin{eqnarray}
i\hbar \left(\frac{d}{dp} + \frac{\beta}{m^2_p} p^2 \frac{d}{dp} + \frac{\beta}{m^2_p} p\right)\varphi_{\pm} = \mp i \varphi_{\pm}\, ,
\end{eqnarray}
belonging to $AC({\mathcal{I}})$. From (\ref{3ab}) we see that the general
solutions are
\begin{eqnarray}
\varphi_{\pm}(p) \ = \  B_{\pm} \frac{e^{\mp m_p{\mbox{\scriptsize{arctanh}}}\left(p \sqrt{|\beta|}/m_p  \right)/\hbar\sqrt{|\beta|}}}{\sqrt{m^2_p  \ - \ p^2 |\beta|}}\, ,
\end{eqnarray}
which are not quadratically integrable on ${\mathcal{I}}$ (and hence they cannot belong to $AC({\mathcal{I}})$) so that the deficiency indices are $(0,0)$. This reconfirms our finding that the $\hat{\mathsf{x}}$ operator is self-adjoint for $\beta <0$.

\section*{B. Connection between wave functions in momentum and position representation}
\label{sb}

\subsection{
Positive $\beta$ case}

Any one-dimensional wave function $|\psi \rangle$ can be written in the momentum basis as
\begin{eqnarray}
|\psi \rangle \ = \ \int_{\mathbb{R}} dp \ \!\langle p | \psi \rangle \ \! |p \rangle\, ,
\end{eqnarray}
which implies that
\begin{eqnarray}
\psi(x)  \ = \ \int_{\mathbb{R}} dp \ \! \psi_p(x) \ \! \tilde{\psi}(p)  \ = \ \int_{\mathbb{R}} dp \ \! \sqrt{\frac{m_p}{\pi}} \beta^{1/4}\frac{e^{ixm_p\arctan\left(p \sqrt{\beta}/m_p  \right)/\hbar\sqrt{\beta}}}{\sqrt{m^2_p  \ + \ p^2 \beta}} \ \! \tilde{\psi}(p)\, ,
\label{B.7c}
\end{eqnarray}
where $\tilde{\psi}(p) = \langle p | \psi \rangle$.  Eq.~(\ref{B.7c}) provides a dictionary between position and momentum representation. Similarly, the inverse transformation reads
\begin{eqnarray}
\tilde{\psi}(p) &=& \sum_{n \in \mathbb{Z}} \ \!  \psi_{(x = n 2\hbar \sqrt{\beta }/m_p)}(p)\ \! {\psi}(x = n 2\hbar \sqrt{\beta }/m_p)\nonumber\\[2mm]
&=&  \sum_{n \in \mathbb{Z}} \ \! \sqrt{\frac{m_p}{\pi}} \beta^{1/4}\frac{e^{-2 in\arctan\left(p \sqrt{\beta}/m_p  \right)}}{\sqrt{m^2_p  \ + \ p^2 \beta}} \ \! {\psi}(x = n 2\hbar \sqrt{\beta }/m_p)\, .
\label{B.8c}
\end{eqnarray}
As a consistency check one can multiply  both sides of Eq.~(\ref{B.7c}) by $\psi_{x}(p')$ and then sum over $n \in \mathbb{Z}$. This yields
\begin{eqnarray}
\tilde{\psi}(p')  \ = \ \sum_{n \in \mathbb{Z}} \ \!  \int_{\mathbb{R}} dp \ \! {\frac{m_p}{\pi}} \sqrt{\beta} \ \! \frac{e^{2 i n[\arctan\left(p \sqrt{\beta}/m_p  \right) - \arctan\left(p' \sqrt{\beta}/m_p  \right)]}}{\sqrt{m^2_p  \ + \ p'^2 \beta}\sqrt{m^2_p  \ + \ p^2 \beta}}\ \! \tilde{\psi}(p)\, .
\label{B9ac}
\end{eqnarray}
By employing the identity
\begin{eqnarray}
\frac{1}{2\pi} \sum_{k \in \mathbb{Z}} e^{i n z} \ = \ \sum_{k \in \mathbb{Z}} \delta(z - 2\pi k)\, ,
\end{eqnarray}
we can rewrite (\ref{B9ac}) as
\begin{eqnarray}
\tilde{\psi}(p')  &=& \sum_{n = \{-1,0,1\}} \ \! \int_{\mathbb{R}} dp \ \! {m_p}\sqrt{\beta} \ \! \frac{\delta\left(\arctan\left(p \sqrt{\beta}/m_p  \right) - \arctan\left(p' \sqrt{\beta}/m_p  \right) - \pi n \right)}{\sqrt{m^2_p  \ + \ p'^2 \beta}\sqrt{m^2_p  \ + \ p^2 \beta}}\ \! \tilde{\psi}(p)\nonumber \\[2mm]
&=& \int_{\mathbb{R}} dp \ \! \delta(p-p') \ \!\tilde{\psi}(p) \ = \  \tilde{\psi}(p')\, ,
\end{eqnarray}
where on the first line we have used the fact that in $\sum_{k \in \mathbb{Z}}$ survive only terms with $n = \{-1,0,1\}$ because $\max_z[ \arctan\left(z\right)] = \pi/2$ (reached for $z\rightarrow \infty$) and $\min_z [\arctan\left(z\right)] = -\pi/2$ (reached for $z\rightarrow -\infty$). On the second line we have realized that contributions from $n=\pm 1$ do not need to be considered as they contribute only for $p' \rightarrow \infty$ and $p \rightarrow -\infty$ or
$p' \rightarrow -\infty$ and $p \rightarrow \infty$, and in these cases  $\tilde{\psi} \rightarrow 0$ (as it is an elements of the $L^2$ Hilbert space).

Similarly, by multiplying Eq.~(\ref{B.8c}) by $\psi_{p}(x')$ and integrating over $p$ we get
\begin{eqnarray}
&&{\psi}(x' = n' 2\hbar \sqrt{\beta }/m_p)\nonumber \\[2mm]
&&\mbox{\hspace{5mm}}= \ \int _{\mathbb{R}} dp \ \! \sqrt{\frac{m_p}{\pi}} \beta^{1/4}\frac{e^{ix'm_p\arctan\left(p \sqrt{\beta}/m_p  \right)/\hbar\sqrt{\beta}}}{\sqrt{m^2_p  \ + \ p^2 \beta}} \ \!
\sum_{n \in \mathbb{Z}} \ \! \sqrt{\frac{m_p}{\pi}} \beta^{1/4}\frac{e^{-2 i n\arctan\left(p \sqrt{\beta}/m_p  \right)}}{\sqrt{m^2_p  \ + \ p^2 \beta}} \ \! {\psi}(x = n 2\hbar \sqrt{\beta }/m_p)\nonumber \\[2mm]
&&\mbox{\hspace{5mm}}= \  \sum_{n \in \mathbb{Z}} \ \! \int _{\mathbb{R}} dp \ \! {\frac{m_p}{\pi}} \sqrt{\beta} \ \! \frac{e^{-2i(n-n')\arctan\left(p \sqrt{\beta}/m_p  \right)}}{m^2_p  \ + \ p^2 \beta}\ \! {\psi}(x = n 2\hbar \sqrt{\beta }/m_p)\nonumber \\[2mm]
&&\mbox{\hspace{5mm}}= \  \sum_{n \in \mathbb{Z}} \ \!  \int_{-\pi}^{\pi} dz \ \! \frac{1}{2\pi} \ \! e^{-i (n-n') z} \ \! {\psi}(x = n 2\hbar \sqrt{\beta }/m_p)\nonumber \\[2mm]
&&\mbox{\hspace{5mm}}= \  \sum_{n \in \mathbb{Z}} \ \! \underbrace{\frac{\sin[(n-n')\pi]}{[(n-n')\pi]}}_{\delta_{nn'}} \ \!  {\psi}(x = n 2\hbar \sqrt{\beta }/m_p) \ = \  {\psi}(x' = n' 2\hbar \sqrt{\beta }/m_p)\, .
\end{eqnarray}
 Unfortunately, as we have already mentioned, positive value of $\beta$ is not compatible with canonical commutation relation {\cdred (2)}.

\subsection{
 Negative $\beta$ case}

In this case the position and momentum representations of a wave function are related via relation
\begin{eqnarray}
\psi(x)\ &=& \ \int_{-m_p/\sqrt{|\beta|}}^{m_p/\sqrt{|\beta|}} \frac{dp}{\sqrt{2\pi \hbar}} \ \! \frac{e^{ixm_p{\mbox{\scriptsize{arctanh}}}\left(p \sqrt{|\beta|}/m_p  \right)/\hbar\sqrt{|\beta|}}}{\sqrt{1   -  p^2 |\beta|/m_p^2}}\ \! \tilde{\psi}(p)\nonumber \\[2mm]
&=& \ \{ z =  m_p\mbox{{arctanh}}\left(p \sqrt{|\beta|}/m_p \right)/\sqrt{|\beta|}\}\nonumber \\[2mm]
&=& \ \int_{\mathbb{R}} \frac{dz}{\sqrt{2\pi \hbar}} \ \! e^{ixz/\hbar} \ \! \frac{\tilde{\psi}(m_p\tanh(z \sqrt{|\beta|}/m_p)/\sqrt{|\beta|})}{\cosh(z \sqrt{|\beta|}/m_p)} \nonumber
 \\[2mm]
&=& \ \int_{\mathbb{R}} \frac{dz}{\sqrt{2\pi \hbar}} \ \! e^{ixz/\hbar} \bar{\psi}(z)   \, ,
\label{B.7cd}
\end{eqnarray}
where $\bar{\psi}(z) = \tilde{\psi}(m_p\tanh(z \sqrt{|\beta|}/m_p)/\sqrt{|\beta|})/ {\cosh(z \sqrt{|\beta|}/m_p)}$. Note that this formula holds only in $D=1$ dimensions. In passing we can easily check that the analogue of  Parseval--Plancherel theorem holds, namely
\begin{eqnarray}
\int_{\mathbb{R}} dx \ \! |\psi(x)|^2 \ = \  \int_{-m_p/\sqrt{|\beta|}}^{m_p/\sqrt{|\beta|}} {d p} \ \! |\tilde{\psi}(p)|^2 \ = \
\int_{\mathbb{R}} {d z} \ \! |\bar{\psi}(z)|^2 \;\;\; \Leftrightarrow \;\;\; |\!| \psi|\!|_2 \ = \ |\!| \tilde{\psi}|\!|_2 \ = \ |\!| \bar{\psi}|\!|_2   \, .
\end{eqnarray}
From the last line in (\ref{B.7cd}) one can also easily deduce that the momentum operator in the position representation has the form
\begin{eqnarray}
\hat{p}^{(x)} \ = \ m_p \tanh\left(-i \hbar \sqrt{|\beta|}/m_p \ \! \frac{d}{dx} \right)/\sqrt{|\beta|}\, .
\end{eqnarray}
It can be easily checked that this operator indeed satisfies the canonical commutation relation {\cdred (2)}.

There is yet another interesting consequence of Eq.~(\ref{B.7cd}), namely one can directly compute from it the corresponding {\em position-space coherent state}. In particular, by using the Tsallis probability amplitude {\cdred (11)} (i.e., momentum-space coherent state) we can write for the corresponding position-space coherent state $\psi_{CS}(x)$ that
\begin{eqnarray}
\psi_{CS}(x) &=& N \int_{\mathbb{R}} \frac{dz}{\sqrt{2\pi \hbar}} \ \! e^{ixz/\hbar}  \frac{\left[m^2_p - m^2_p (\tanh(z \sqrt{|\beta|}/m_p))^2 \right]^{m_p^2/(2|\beta| \gamma \hbar) - 1/2}}{\cosh\left(z \sqrt{|\beta|}/m_p \right)}\nonumber \\[2mm]
&=& N m_p^{m_p^2/(|\beta| \gamma \hbar)-1} \int_{\mathbb{R}} \frac{dz}{\sqrt{2\pi \hbar}} \ \! e^{ixz/\hbar} \ \! {\cosh\left(z \sqrt{|\beta|}/m_p \right)^{-m_p^2/(|\beta| \gamma \hbar)}}\nonumber \\[2mm]
&=& \tilde{N }
\left|\Gamma\left(\frac{m^2_p/(|\beta| \gamma \hbar)}{2} + i \frac{xm_p}{2\sqrt{|\beta|}\hbar}\right)\right|^2 \nonumber \\[2mm]
&=&  \sqrt{\frac{m_p \ \!\Gamma(2 m^2_p/(|\beta| \gamma \hbar))}{4\pi \sqrt{|\beta|} \hbar  \ \! \Gamma^4(m^2_p/(|\beta| \gamma \hbar))}}\; \left|\Gamma\left(\frac{m^2_p/(|\beta| \gamma \hbar)}{2} + i \frac{xm_p}{2\sqrt{|\beta|}\hbar}\right)\right|^2 \, .
\label{CS-x}
\end{eqnarray}
In passage from 2nd to 3rd line we used the fact that the 2nd line represents a Fourier transform (characteristic function)
of {\em generalized logistic density} of type III~\cite{distributions}, cf. also Ref.~\cite{Ober}. Alternatively one can obtain the result (\ref{CS-x}) from the Ramanujan formula~\cite{Ramanujan}
\begin{eqnarray}
\int_{-\infty}^{\infty} e^{-i\xi s} \ \! \left|\Gamma(a + is) \right|^2 \ \! ds \ = \ \sqrt{\pi} \Gamma(a)\Gamma(a+ 1/2) \left[ \cosh(\xi/2)\right]^{-2a} \ = \ \frac{2\pi}{2^{2a}}  \Gamma(2a) \left[ \cosh(\xi/2)\right]^{-2a}  \, ,
\end{eqnarray}
that is valid for $a\in (-1,0)\cup (0,\infty)$. The normalization factor on the last line of (\ref{CS-x}) was obtained by using the Mellin--Barnes beta integral (cf. Ref.~\cite{Ober})
\begin{eqnarray}
\int_{-\infty}^{\infty} \ \! \left|\Gamma(a + ib s) \right|^4 \ \! ds \ = \ \frac{2 \pi}{b} \frac{\Gamma^4(2a)}{\Gamma(4a)}\, .
\label{19.SM.aa}
\end{eqnarray}
In passing we note that the state $\psi_{CS}(x)$ is even-parity state (as required) and, in addition,  it belongs to the Schwartz class, i.e., it decays rapidly at infinity along with all derivatives.

For consistency we can now check that $\psi_{CS}(x)$ from Eq.~(\ref{CS-x}) provides a correct positional variance, which together with the momentum variance deduced from  $\tilde{\psi}_{CS}(p)$ (cf. {\cdred Eq.~(19)}) saturates the GUP {\cdred (1)}.
To this end we use the formula for the Fourier transform of  $\left|\Gamma(a + ib s) \right|^4$ (see, \cite{Ober}, Eq.~(274), p. 46) to show that
\begin{eqnarray}
\frac{b \Gamma(4a)}{2\pi \Gamma^4(2a)}\int_{-\infty}^{\infty} \ \! s^2 \ \! \left|\Gamma(a + ib s) \right|^4 \ \! ds \ = \ \frac{a^2}{b^2(1+4a)}\, .
\end{eqnarray}
If we now use from (\ref{19.SM.aa}) that $a = m_p^2/(2|\beta|\gamma \hbar)$ and $b = m_p/(2\sqrt{|\beta|}\hbar)$ we get
\begin{eqnarray}
(\Delta x)^2_{CS} \ = \ \int_{-\infty}^{\infty} \ \! x^2 \ \! \psi_{CS}(x) \ \! dx \ = \ \frac{\hbar m_p^2}{2m_p^2 \gamma + \hbar |\beta| \gamma^2} \ = \ \frac{(\Delta p)_{CS}^2}{\gamma^2}\, ,
\label{SM.40.cc}
 \end{eqnarray}
where the last identity results from {\cdred(15)}. This is equivalent to the saturated GUP.

\section*{C. Generalization of Beckner--Babenko's theorem} \label{sc}
\subsection{Beckner--Babebko's theorem for Pontryagin class of transformations}

Important for Fourier-transform based entropic uncertainty relations is the following theorem~\cite{Beckner1975,Babenko1962}:

\begin{theorem} [Beckner--Babebko's theorem]
Let
\[
f^{(2)}({{\boldsymbol{x}}}) \equiv
\hat{f}^{(1)}({{\boldsymbol{x}}})= \int_{\mathbb{R}^{D}}e^{2\pi
i{{\boldsymbol{x}} }.{{\boldsymbol{y}}}}\
f^{(1)}({{\boldsymbol{y}}})\ d{{\boldsymbol{y}}}\,,
\]
then for $p \in [1,2]$ we have
\begin{eqnarray}
|\!|\hat{f}|\!|_{p^{\prime}}\ \leq\
\frac{|p^{D/2}|^{1/p}}{|(p^{\prime} )^{D/2}|^{1/p^{\prime}}}\
|\!|f|\!|_{p}\, , \label{4.1}
\end{eqnarray}
or, equivalently
\begin{eqnarray*}
\;|(p^{\prime})^{D/2}|^{1/p^{\prime}}|\!|f^{(2)}|\!|_{p^{\prime}} \
\leq\ |p^{D/2}|^{1/p}|\!|f^{(1)}|\!|_{p}\, .
\end{eqnarray*}
Here, $p$ and $p^{\prime}$ are the usual H\"{o}lder
conjugates (i.e., $p' \in [2,\infty)$). For any  $X \in
\ell^{p}({\mathbb{R}^{D}})$ the $p$-norm $|\!|X|\!|_{p}$ is defined as
\[
|\!|X|\!|_{p} = \left(\int_{\mathbb{R}^{D}}
|X({\boldsymbol{y}})|^{p} \ d{{\boldsymbol{y}}}\right)^{1/p}.
\]
Due to symmetricity of the Fourier transform also the reverse inequality holds:
\begin{eqnarray}
|\!|{f}|\!|_{p^{\prime}}\ \leq\
\frac{|p^{D/2}|^{1/p}}{|(p^{\prime} )^{D/2}|^{1/p^{\prime}}}\
|\!|\hat{f}|\!|_{p}\, . \label{4.1b}
\end{eqnarray}
\end{theorem}

Lieb~\cite{Lieb1990} proved that the inequality (\ref{4.1}) is
saturated (jointly for all $p$) only for Gaussian functions. This relation directly implies entropy-power uncertainty relation for
Pontryagin class of transformations (cf. Section~\ref{sf}).

\vspace{3mm}

B-B theorem can be directly generalized to our situation (\ref{B.7cd}). This will be discussed in Section~\ref{GUP}.
%

\section*{D. Entropy powers} \label{sd}
\subsection{Entropy powers based on Gaussian distribution}

\noindent {\em R\'{e}nyi entropy power}~~R\'{e}nyi entropy power $N_p^R(\mathcal{X})$ is defined as the
solution of the equation
\begin{eqnarray}
{\mathcal{S}}_q^R\left( {\mathcal{X}} \right)
\ = \ {\mathcal{S}}_q^R\left(\sqrt{N_q^R(\mathcal{X})}\cdot
{\mathcal{Z}}^{_G}\right)\,
,\label{3.1.0k}
\end{eqnarray}
where $\{{\mathcal{Z}}_{i}^{_G}\}$ represents a {\em Gaussian random vector} with zero mean
and unit covariance matrix. So, $N_p^R(\mathcal{X})$ denotes the variance of a would be Gaussian distribution that
has the same R\'{e}nyi information content as the random vector $\{\mathcal{X}_i\}$ described by the PDF ${\mathcal{F}}({\boldsymbol{x}})$.
Expression (\ref{3.1.0k}) was studied in~\cite{JD:16,JDJ,Gardner02} where it was shown
that the only  class of solutions of (\ref{3.1.0k}) is
\begin{eqnarray}
N_q^R(\mathcal{X})
\ &=& \ \frac{1}{2\pi} p^{-q'/q}
\exp\left(\frac{2}{D} \ \!{\mathcal{S}}_q^R({\mathcal{X}})\right),
\label{3.1.0e}
\end{eqnarray}
with $1/p + 1/p'= 1$ and $p\in {\mathbb{R}}^{+}$.
In addition, when $p \rightarrow 1_+$ one has
$N_p^R(\mathcal{X}) \rightarrow N(\mathcal{X})$, where $N(\mathcal{X})$ is the conventional Shannon entropy power~\cite{Shannon}.

\vspace{3mm}

\noindent {\em Tsallis entropy power}~~Tsallis entropy power $N_p^T(\mathcal{X})$ is defined as the
solution of the equation
\begin{eqnarray}
{\mathcal{S}}_q^T\left( {\mathcal{X}} \right)
\ = \ {\mathcal{S}}_q^T\left(\sqrt{N_q^T(\mathcal{X})}\cdot
{\mathcal{Z}}^{_G}\right)\,
.\label{3.1.0l}
\end{eqnarray}
Corresponding entropy power has not been studied in the literature yet but it can easily be derived by observing that
the following scaling property for differential Tsallis entropy holds, namely
\begin{eqnarray}
{\mathcal{S}}_q^T(a \mathcal{X}) \ = \   {\mathcal{S}}_q^T(\mathcal{X}) \ \oplus_q \ \ln_q |a|^D\, ,
\label{14.bb}
\end{eqnarray}
where $a \in \mathbb{R}$ and the $q$-deformed ``sum'' and logarithm are defined as~\cite{tsallis-book}: $x \oplus_q  y  =  x  +  y  +  (1-q) x y$ and
$\ln_q x  =  (x^{1-q }- 1)/(1-q)$, respectively.
Relation (\ref{14.bb}) follows directly from the following chain of identities
\begin{eqnarray}
{\mathcal{S}}_q^T(a \mathcal{X}) \ &=& \  \frac{1}{1-q} \left[\int d^D {\bf y}\left(\int d^D {\bf x} \ \! \delta({\bf y} - a {\bf x}) {\mathcal{F}}({\boldsymbol{x}}) \right)^{\!q} -1 \right] \ = \ \frac{1}{1-q} \left[|a|^{D(1-q)}\ \! \int d^D {\bf y} \ \! {\mathcal{F}}^q({\boldsymbol{y}}) -1  \right]\nonumber \\[2mm]
&=& \ |a|^{D(1-q)}\ \! \left({\mathcal{S}}_q^T(\mathcal{X}) \ + \ \frac{1}{1-q} \right) \ - \ \frac{1}{1-q} \ = \ |a|^{D(1-q)}\ \!{\mathcal{S}}_q^T(\mathcal{X}) \ + \ \ln_q |a|^D
\nonumber \\[2mm]
&=& \ \left[(1-q) \ln_q |a|^D \ + \  1\right]{\mathcal{S}}_q^T(\mathcal{X}) \ + \ \ln_q |a|^D
\ = \ {\mathcal{S}}_q^T(\mathcal{X}) \ \oplus_q \ \ln_q |a|^D\, .
\label{scaling}
\end{eqnarray}
We can further use the simple fact that
\begin{eqnarray}
{\mathcal{S}}_q^T({\mathcal{Z}}_{G}) \ = \  \ln_q(2\pi q^{q'/q})^{D/2}\, .
\label{15.bc}
\end{eqnarray}
Here $p$ and $p'$ are H\"{o}lder's double, i.e., $1/q + 1/q' =1$. Combining (\ref{3.1.0l}), (\ref{14.bb}) and (\ref{15.bc}) we get that
\begin{eqnarray}
{\mathcal{S}}_q^T(\mathcal{X}) \ = \ \ln_q(2\pi q^{q'/q})^{D/2} \ \oplus_q \ \ln_q (N_q^T)^{D/2} \ = \ \ln_q(2\pi q^{q'/q}N_q^T)^{D/2}\, ,
\label{15.bcd}
\end{eqnarray}
where we have used the sum rule from the $q$-deformed calculus: $\ln_q x \oplus_q  \ln_q y = \ln_q xy$. Equation (\ref{15.bcd}) can be resolved for $N_p^T$ when we employ the $q$-exponential, i.e., $e_q^x = [1 + (1-q)x]^{1/(1-q)}$, which among others satisfies the relation $e_q^{\ln_q x} = \ln_q (e_q^x) = x$. With this we have that
\begin{eqnarray}
N_q^T(\mathcal{X}) \ = \ \frac{1}{2\pi} q^{-q'/q} \left[\exp_q \left({\mathcal{S}}_q^T(\mathcal{X})\right)\right]^{2/D} \ = \ \frac{1}{2\pi} q^{-q'/q} \exp_{1 - (1-q)D/2} \left(\frac{2}{D}\ \!{\mathcal{S}}_q^T(\mathcal{X})\right)\, .
\end{eqnarray}
In addition, when $p \rightarrow 1_+$ one has
\begin{eqnarray}
\lim_{q\rightarrow 1 }N_q^T(\mathcal{X}) \ = \ \frac{1}{2\pi e} \exp\left( \frac{2}{D} \mathcal{H}(\mathcal{X})\right) \ = \ N(\mathcal{X})\, ,
\end{eqnarray}
where $N(\mathcal{X})$ is the conventional Shannon entropy power and $\mathcal{H}(\mathcal{X})$ is Shannon entropy~\cite{Shannon}.

In connection with Tsallis entropy power we might notice one interesting fact, namely by starting with R\'{e}nyi entropy power we have
\begin{eqnarray}
N_q^R(\mathcal{X})
\ &=& \ \frac{1}{2\pi} q^{-q'/q}
\exp\left(\frac{2}{D} \ \!{\mathcal{S}}_q^R({\mathcal{X}})\right) \ = \ \frac{1}{2\pi} q^{-q'/q} \left(\int d^D{\bf x} \ \! {\mathcal{F}}^q({\boldsymbol{x}})  \right)^{2/(D(1-q))} \nonumber \\[2mm]
&=& \ \frac{1}{2\pi} q^{-q'/q} \left[e_q^{{\mathcal{S}}_q^T(\mathcal{X})}\right]^{2/D} \ = \ N_q^T(\mathcal{X})\, .
\label{24aa}
\end{eqnarray}
Here we have used the obvious identity
\begin{eqnarray}
\left(\int d^D{\bf x} \ \! {\mathcal{F}}^q({\boldsymbol{x}})  \right)^{{1}/{(1-q)}} \ = \ \left[(1-q){\mathcal{S}}_q^T({\mathcal{X}}) \ + \ 1   \right]^{1/(1-q)} \ = \ e_q^{{\mathcal{S}}_q^T(\mathcal{X})}\, .
\end{eqnarray}
So, we have obtained that R\'{e}nyi and Tsallis entropy powers match each other.

\subsection{Entropy powers based on Tsallis entropy}

When dealing with GUP  that is saturated by Tsallis probability amplitude states, it is more convenient to work with entropy powers based on
Tsallis distribution {\cdred (17)}. In this connection we can again formally define the ensuing entropy powers for both R\'{e}nyi and Tsallis entropies as solutions of the equations
\begin{eqnarray}
&&{\mathcal{S}}_q^R\left( {\mathcal{X}} \right)
\ = \ {\mathcal{S}}_q^R\left(\sqrt{M_q^R(\mathcal{X})}\cdot
{\mathcal{Z}}^{T}\right)\, ,\nonumber \\[2mm]
&&
{\mathcal{S}}_q^T
\left( {\mathcal{X}} \right)
\ = \ {\mathcal{S}}_q^T\left(\sqrt{M_q^T(\mathcal{X})}\cdot
{\mathcal{Z}}^{T}\right)\, ,
\label{3.1.0kl}
\end{eqnarray}
where $\{{\mathcal{Z}}_{i}^{T}\}$ represents a {\em Tsallis random vector} with zero mean
and unit covariance matrix. Such a vector is distributed with respect to the $q$-Gaussian probability density function that extremizes ${\mathcal{S}}_q^T$ (and hence also ${\mathcal{S}}_q^R$).
To solve Eqs.~(\ref{3.1.0kl}) we use the scaling relations for R\'{e}nyi and Tsallis entropies [see, Eq.~(\ref{scaling})], namely
\begin{eqnarray}
{\mathcal{S}}_q^R(a \mathcal{X}) \ &=& \ {\mathcal{S}}_q^R(\mathcal{X}) \ + \ D \ln |a|\, , \nonumber \\[2mm]
{\mathcal{S}}_q^T(a \mathcal{X}) \ &=& \ {\mathcal{S}}_q^T(\mathcal{X}) \ \oplus_q \ \ln_q |a|^D\, .
\end{eqnarray}
In the next step we need to know ${\mathcal{S}}_q^R\left({\mathcal{Z}}^{T}\right)$ and ${\mathcal{S}}_q^T\left({\mathcal{Z}}^{T}\right)$.
To this end we first realise that the $q$-Gaussian distribution extemizing ${\mathcal{S}}_{2-q}^T$ with zero mean
and unit covariance matrix has the form~\cite{tsallis-book}
\begin{eqnarray}
&&\mathcal{F}(q,{\bf p}) \ = \ {N}_{_{\lessgtr}}(q)\left[ 1 \ - \ \mathfrak{b}{(1-q)}\ \!{\bf p}^2\right]_+^{1/(1-q)}\;
\; \;\; \mbox{with} \;\;\; \; \mathfrak{b} \ = \ \frac{1}{ \left[2(2-q) -D (q-1)\right]}\, ,
\end{eqnarray}
Note that $\mathfrak{b} > 0$ for all $q< (4+D)/(2+D)$. For $q\geq (4+D)/(2+D)$ the variance is not defined.
The normalization factor $N_{_{\lessgtr}}(q)$ is  such that for $q<1$
\begin{eqnarray}
N_<(q) \ &=& \ \left(1+ \frac{D}{2}(1-q) \right) \frac{1}{\pi^{D/2}}  \left(\mathfrak{b}{(1-q)}\right)^{D/2}  \ \! \frac{\Gamma\left(\frac{1}{1-q}  + \frac{D}{2}   \right)}{\Gamma\left(\frac{1}{1-q } \right) } \nonumber \\[2mm]
&=& \ \frac{1}{\pi^{D/2}}  \left(\mathfrak{b}{(1-q)}\right)^{D/2}  \ \! \frac{\Gamma\left(\frac{2-q}{1-q}  + \frac{D}{2}   \right)}{\Gamma\left(\frac{2-q}{1-q } \right) }\, ,
\end{eqnarray}
and for $1<q<(4+D)/(2+D)$
\begin{eqnarray}
N_>(q) \ = \ \frac{1}{\pi^{D/2}} \left(\mathfrak{b}{(q-1)}\right)^{D/2} \ \! \frac{\Gamma\left(\frac{1}{q-1}  \right)}{\Gamma\left(\frac{1}{q-1 } - \frac{D}{2}  \right) }  \, .
\end{eqnarray}
Since $\mathcal{F}(2-q,{\bf p})$ extremises ${\mathcal{S}}_{q}^T$  (or equivalently $\mathcal{F}(q,{\bf p})$ extremises ${\mathcal{S}}_{2-q}^T$), we can write
\begin{eqnarray}
\int d^D{\bf p} \ \! [\mathcal{F}(2-q,{\bf p})]^q \ \! &=& \ \! N^q_{_{\lessgtr}}(2-q) \int d^D{\bf p} \left[ 1  \ - \ \mathfrak{b}(q-1) {{\bf p}^2}\right]^{q/(q-1)}_+\nonumber \\[2mm]
&=& N^q_{_{\lessgtr}}(2-q) \Omega_D \int_{0}^{\infty} d z \ \! z^{D-1} \left[ 1  \ - \ {\mathfrak{b}(q-1)} {{z}^2}\right]^{q/(q-1)}_+\, .
\label{SM.58cc}
\end{eqnarray}
Let us now discuss a bit more the situations with ${\mathfrak{b}(q-1)} < 0 $ that is relevant in the main text.

When ${\mathfrak{b} (q-1)} < 0$ [i.e. when $q< 1$] we work with $N_>(2-q)$ and Eq.~(\ref{SM.58cc}) turns to
\begin{eqnarray}
&&N^q_{>}(2-q) \frac{\pi^{D/2}}{\Gamma(D/2)} \left(\frac{1}{\mathfrak{b}(1-q)} \right)^{D/2}\ \!\int_{0}^{\infty} d x \ \! x^{D/2-1}\left[ 1  \ + \ x\right]^{q/(q-1)} \nonumber \\[2mm]
&&\hspace{1cm} = \ N^q_{>}(2-q) \frac{\pi^{D/2}}{\Gamma(D/2)} \left(\frac{1}{\mathfrak{b}(1-q)} \right)^{D/2}\ \!
B\!\left(\frac{D}{2}, \frac{q}{1-q} -\frac{D}{2}\right)\nonumber \\[2mm]
&&\hspace{1cm} = \ N^q_{>}(2-q)  \left(\frac{\pi}{\mathfrak{b}(1-q)} \right)^{D/2}\ \! \frac{\Gamma\left(\frac{q}{1-q} -\frac{D}{2}\right)}{\Gamma\left(\frac{q}{1-q}\right)})\nonumber \\[2mm]
&&\hspace{1cm} = \
 \left(\frac{\mathfrak{b}(1-q)}{\pi}  \right)^{D/2(q-1)}  \frac{\Gamma\left(\frac{q}{1-q} -\frac{D}{2}\right)}{\Gamma\left(\frac{q}{1-q}\right)} \left[ \frac{\Gamma\left(\frac{1}{1-q} \right)}{\Gamma\left(\frac{1}{1-q} -\frac{D}{2}\right)} \right]^q
\nonumber \\[2mm]
&&\hspace{1cm} = \ \left[\left(\frac{\mathfrak{b}(1-q)}{\pi}  \right)^{D/2} \frac{\Gamma\left(\frac{1}{1-q} \right)}{\Gamma\left(\frac{1}{1-q} -\frac{D}{2}\right)}\right]^{q-1} \left(1- \frac{D}{2q}(1-q)\right)^{-1}.
\label{ap.26}
\end{eqnarray}
%
With (\ref{ap.26}) we get that
\begin{eqnarray}
&&{\mathcal{S}}_q^R\left({\mathcal{Z}}^{T}\right)
 \ = \ \log \left[ \left(\frac{\pi}{\mathfrak{b}(1-q)}  \right)^{D/2} \ \! \frac{\Gamma\left(\frac{1}{1-q}-\frac{D}{2} \right)}{\Gamma\left(\frac{1}{1-q} \right)} \ \!  \left(1- \frac{D}{2q}(1-q)\right)^{1/(q-1)}\right]\, ,
\nonumber \\[2mm]
&&{\mathcal{S}}_q^T\left({\mathcal{Z}}^{T}\right) \ = \ \ln_q \left[\left(\frac{\pi}{\mathfrak{b}(1-q)}  \right)^{D/2} \ \! \frac{\Gamma\left(\frac{1}{1-q}-\frac{D}{2} \right)}{\Gamma\left(\frac{1}{1-q} \right)} \ \!  \left(1- \frac{D}{2q}(1-q)\right)^{1/(q-1)} \right]\, ,
\end{eqnarray}
and consequently the ensuing entropy powers are
\begin{eqnarray}
&&M_q^R(\mathcal{X}) \ = \  \exp\left[\frac{2}{D} \ \!\left({\mathcal{S}}_q^R(\mathcal{X}) - {\mathcal{S}}_q^R({\mathcal{Z}}^{T}) \right)\right] \ = \ A \ \! \exp\left({\frac{2}{D} \ \!{\mathcal{S}}_q^R(\mathcal{X})}\right) \nonumber \\[2mm]
&&M_q^T(\mathcal{X}) \ = \ A \ \! \left[\exp_q({\mathcal{S}}_q^T(\mathcal{X}))\right]^{2/D} \ = \ A \ \! \exp_{1-(1-q)D/2}\left({\frac{2}{D} \ \!{\mathcal{S}}_q^T(\mathcal{X})}\right)\, ,
\label{SM.61cc}
\end{eqnarray}
where
\begin{eqnarray}
A \ = \ \left[\left(\frac{\pi}{\mathfrak{b}(1-q)}  \right)^{D/2} \ \! \frac{\Gamma\left(\frac{1}{1-q}-\frac{D}{2} \right)}{\Gamma\left(\frac{1}{1-q} \right)} \ \!  \left(1- \frac{D}{2q}(1-q)\right)^{1/(q-1)}  \right]^{-2/D}\, .
\end{eqnarray}
As a consistency check we can take limit $q \rightarrow 1$. Indeed, by realizing that in the present case  $\sigma^2 =1$ implies  $\mathfrak{b} = [2q - D(1-q)]^{-1}$, we get
\begin{eqnarray}
\lim_{q\rightarrow 1 }M_q^R(\mathcal{X}) \ = \ \lim_{q\rightarrow 1 }M_q^T(\mathcal{X})  \ = \ \frac{1}{2\pi e} \exp\left( \frac{2}{D} \mathcal{H}(\mathcal{X})\right) \ = \ N(\mathcal{X})\, ,
\end{eqnarray}
where $N(\mathcal{X})$ is the conventional Shannon entropy power. Finally we can again check that
\begin{eqnarray}
M_q^R(\mathcal{X}) \ = \ M_q^T(\mathcal{X})\, .
\end{eqnarray}

\section*{F. Entropy power uncertainty relations  \label{sf}}
\subsection{Entropy power uncertainty relations for Pontryagin class of transformations}

In conventional QM  the $\bf{x}$ and $\bf{p}$-representation wave functions $\psi(\bf{x})$ and $\hat{\psi}(\bf{p})$, respectively are related via Fourier transform relations
\begin{eqnarray}
&&\psi({\bf{x}})  \ = \ \int_{\mathbb{R}^D} e^{i {\bf p}\cdot {\bf x}/\hbar} \  \! \hat{\psi}({\bf{p}})\ \!  \frac{d{\bf p}}{(2\pi \hbar)^{D/2}}\, ,\nonumber \\[2mm]
&&\hat{\psi}({\bf{p}})  \ = \ \int_{\mathbb{R}^D} e^{-i {\bf p}\cdot {\bf x}/\hbar} \  \! {\psi}({\bf{x}})\ \!  \frac{d{\bf x}}{(2\pi \hbar)^{D/2}}\, . \label{V.1.a}
\end{eqnarray}
Plancherel (or Riesz--Fischer) equality  then implies that $|\!|{\psi}|\!|_2 = |\!|
\hat{\psi}|\!|_{2} = 1$. Let us define new functions, namely
\begin{eqnarray}
&&f^{(2)}({\bf{x}}) \ = \ (2\pi \hbar)^{D/4}\psi(\sqrt{2\pi\hbar}\ \! {\bf{x}})\, , \nonumber \\[2mm]
&&f^{(1)}({\bf{p}}) \ = \ (2\pi \hbar)^{D/4}\hat{\psi}(\sqrt{2\pi\hbar}\ \! {\bf{p}}) \, .
\label{6.2.a}
\end{eqnarray}
The factor $(2\pi \hbar)^{D/4}$ ensures that also the new functions are normalized (in sense of $|\!|\ldots|\!|_2$) to unity.
With these we will have the same structure of the Fourier transform  as in the Beckner--Babenko inequality. Beckner--Babenko inequality (\ref{4.1b})
can be then rewritten as
\begin{eqnarray}
\left[\left(\frac{q'}{2\pi \hbar} \right)^D\right]^{1/q'} |\!| |\psi|^2 |\!|_{q'/2}  \ \leq \ \left[\left(\frac{q}{2\pi \hbar} \right)^D\right]^{1/q} |\!| |\hat{\psi}|^2 |\!|_{q/2}\, .
\end{eqnarray}
This is equivalent to
\begin{eqnarray}
\left[\left(\frac{q'}{2\pi \hbar} \right)^D\right]^{1/q'} \exp\left[\frac{2(1-q'/2)}{q'} \ \!{\mathcal{S}}_{q'/2}^R\left( |\psi|^2 \right)\right] \ \leq \ \left[\left(\frac{q}{2\pi \hbar} \right)^D\right]^{1/q} \exp\left[\frac{2(1-q/2)}{q} \ \!{\mathcal{S}}_{q/2}^R\left( |\hat{\psi}|^2 \right)\right].
\end{eqnarray}
Now we take power $q/(D(1-q/2))$ of both left and right side and use the fact that $2/q - 1 = 1- 2/q'$, whis gives
\begin{eqnarray}
\left(\frac{q}{2\pi \hbar}\right)^{1/(1-q/2)} \exp\left[\frac{2}{D} \ \!{\mathcal{S}}_{q/2}^R\left( |\hat{\psi}|^2 \right)\right] \left(\frac{q'}{2\pi \hbar}\right)^{1/(1-q'/2)} \exp\left[\frac{2}{D} \ \!{\mathcal{S}}_{q'/2}^R\left( |\psi|^2 \right)\right]     \  \geq \ 1\, .
\end{eqnarray}
This is identical to (use that for H\"{o}lder double one has $1/(1-q/2) + 1/(1-q'/2) =2$)
\begin{eqnarray}
\underbrace{\frac{1}{2\pi} \left(\frac{q}{2}\right)^{1/(1-q/2)} \exp\left[\frac{2}{D} \ \!{\mathcal{S}}_{q/2}^R\left( |\hat{\psi}|^2 \right)\right]}_{N_{q/2}^R(|\hat{\psi}|^2)}
\underbrace{\frac{1}{2\pi} \left(\frac{q'}{2}\right)^{1/(1-q'/2)} \exp\left[\frac{2}{D} \ \!{\mathcal{S}}_{q'/2}^R\left( |\psi|^2 \right)\right]}_{N_{q'/2}^R(|{\psi}|^2)}     \  \geq \ \frac{\hbar^2}{4}\, .
\end{eqnarray}
By using (\ref{24aa}) this is equivalent to
\begin{eqnarray}
N_{q/2}^T(|\hat{\psi}|^2)N_{q'/2}^T(|{\psi}|^2)  \  \geq \ \frac{\hbar^2}{4}\, .
\end{eqnarray}
This result can be generalized to class of Pontryagin dual wave functions.

\subsection{Entropy power inequalities for GUP transformations \label{GUP}}

Using the relation (\ref{B.7cd}) together with (\ref{6.2.a}) we can write
\begin{eqnarray}
 \left[\left(\frac{q'}{2\pi \hbar} \right)^D\right]^{1/q} |\!| |\bar{\psi}|^2 |\!|_{q'/2}  \ &\leq& \
\left[\left(\frac{q}{2\pi \hbar} \right)^D\right]^{1/q'} |\!| |\psi|^2 |\!|_{q/2}
\, ,
\label{45cc}
\end{eqnarray}
where $q' \in [2,\infty)$ while $q \in [1,2]$.
It can be checked numerically that this is indeed saturated for coherent states
$|\tilde{\psi}|^2_{CS}(p)=
q_T(p|2-q'/2,b)$ [given by {\cdred (17)}] and associated
$\psi_{CS}(x)$ [given by (\ref{CS-x})] with the non-extensivity index $2-q/2$. Analytical proof can be readily done, e.g., for cases $q=1$ and $q=2$.
We can further rewrite (\ref{45cc}) as
\begin{eqnarray}
\left\{{\frac{1}{2\pi} \left(\frac{q}{2}\right)^{1/(1-q/2)} \exp\left[\frac{2}{D} \ \!{\mathcal{S}}_{q/2}^R\left( |\psi|^2 \right)\right]}\right\}
\left\{{\frac{1}{2\pi} \left(\frac{q'}{2}\right)^{1/(1-q'/2)} \exp\left[\frac{2}{D} \ \!{\mathcal{S}}_{q'/2}^R\left( |\bar{\psi}|^2 \right)\right]}\right\}    \  \geq \ \frac{\hbar^2}{4}\, .
\label{SM.77.cc}
\end{eqnarray}
Since the relevant wave function is not $\bar{\psi}$ but $\tilde{\psi}$, cf. Eq.~(\ref{B.7cd}), we need to reformulate (\ref{SM.77.cc}) in terms of $\tilde{\psi}$. This can be done by realizing that
\begin{eqnarray}
&&\mbox{\hspace{-10mm}}\left\{{\frac{1}{2\pi} \left(\frac{q'}{2}\right)^{1/(1-q'/2)} \exp\left[\frac{2}{D} \ \!{\mathcal{S}}_{q'/2}^R\left( |\bar{\psi}|^2 \right)\right]}\right\} \nonumber \\[2mm]
&&= \ \frac{1}{2\pi} \left(\frac{q'}{2}\right)^{1/(1-q'/2)} \left[
\int_{m_p/\sqrt{\beta}}^{m_p/\sqrt{\beta}} dp \ \! \left(|\tilde{\psi}(p)|^2   \right)^{q'/2}\left( 1 - \frac{p^2 \beta}{m_p^2} \right)^{q'/2-1}\right]^{2/(1-q'/2)} \nonumber \\[2mm]
&& = \ \left\{{\frac{1}{2\pi} \left(\frac{q'}{2}\right)^{1/(1-q'/2)} \exp\left[\frac{2}{D} \ \!{\mathcal{S}}_{q'/2}^R\left( |\tilde{\psi}|^2 \right)\right]}\right\} \left\langle  \left( 1 - \frac{p^2 \beta}{m_p^2} \right)^{q'/2-1}  \right\rangle_{q'/2}^{2/(1-q'/2)}\, ,
\label{SM.78.cca}
\end{eqnarray}
where $\langle \ldots \rangle_{x}$ denotes the average value with respect to {\em escort distribution} of order $z$ associated with $f \equiv |\tilde{\psi}|^2$, namely
\begin{eqnarray}
f(p) \mapsto \mathcal{F}_{z}(p) \ = \ \frac{f^{z}(p)}{\int {dp} \ \! f^{z}(p) } \, .
\end{eqnarray}
This allows to write (\ref{SM.77.cc}) as
\begin{eqnarray}
N_{q/2}^T(|{\psi}|^2)N_{q'/2}^T(|\tilde{\psi}|^2) \left\langle  \left( 1 - \frac{p^2 \beta}{m_p^2} \right)^{q'/2-1}  \right\rangle_{q'/2}^{2/(1-q'/2)} \  \geq \ \frac{\hbar^2}{4}\, .
\label{SM.79.ccb}
\end{eqnarray}
Now, without loss of generality, we may assume that  the same value of $\langle \ldots \rangle_{q'/2}$ in (\ref{SM.79.ccb}) could be computed also with distribution $|\tilde{\psi}|^2_{CS}(p)=
q_T(p|2-q'/2,\tilde{b})$ for some unknown parameter $\tilde{b}$, so that also $\beta = bm_p (q'/2-1)/2 \neq \tilde{\beta} = \tilde{b}m_p (q'/2-1)/2$. For such $|\tilde{\psi}|^2_{CS}(p)$ and ensuing  $|{\psi}|^2_{CS}(x)$ one would have
\begin{eqnarray}
N_{q/2}^T(|{\psi}|^2_{CS})N_{q'/2}^T(|\tilde{\psi}|^2_{CS}) \left\langle  \left( 1 - \frac{p^2 \tilde{\beta}}{m_p^2} \right)^{q'/2-1}  \right\rangle_{q'/2}^{2/(1-q'/2)}\ = \ \frac{\hbar^2}{4}\, .
\label{SM.80.ccc}
\end{eqnarray}
By dividing (\ref{SM.79.ccb}) by (\ref{SM.80.ccc}) we obtain
\begin{eqnarray}
M_{q/2}^T(|{\psi}|^2)M_{q'/2}^T(|\tilde{\psi}|^2) \left[\sigma_p^2(\tilde{b},2-q'/2) \sigma_x^2(\tilde{b},2-q/2)   \right]^{-1}  \  \geq \ 1\, ,
\label{SM.81.aac}
\end{eqnarray}
where $\sigma_p^2(\tilde{b},2-q'/2)$ denotes the variance of $|\tilde{\psi}|^2_{CS}(p)=
q_T(p|2-q'/2,\tilde{b})$ and similarly  $\sigma_x^2(\tilde{b},2-q'/2)$ is the variance of the associated $|{\psi}|^2_{CS}(x)$. In deriving (\ref{SM.81.aac}) we used the fact that
\begin{eqnarray}
M_{q'/2}^T(|\tilde{\psi}|^2) \ &=& \ \exp\left[\frac{2}{D}\left(\mathcal{S}_{q'/2}^R(|\tilde{\psi}|^2) -    \mathcal{S}_{q'/2}^R(|\tilde{\psi}|^2_{CS,1\!\!\rm{I}}) \right)\right] \nonumber \\[2mm]
&=& \ \exp\left[\frac{2}{D}\left(\mathcal{S}_{q'/2}^R(|\tilde{\psi}|^2) -    \mathcal{S}_{q'/2}^R(|\tilde{\psi}|^2_{CS}) \right)\right]\left[\sigma_p^2(\tilde{b},2-q'/2)\right]\, .
\end{eqnarray}
Here $\mathcal{S}_{q'/2}^R(|\tilde{\psi}|^2_{CS,1\!\!\rm{I}})$ denotes the R\'{e}nyi entropy of the coherent state distribution with unit variance. Similar relation holds also for $|{\psi}|^2_{CS}(x)$. At this stage we use the formula (\ref{SM.40.cc}) and {\cdred (19)}
to write (\ref{SM.81.aac}) as
\begin{eqnarray}
M_{q/2}^T(|{\psi}|^2)M_{q'/2}^T(|\tilde{\psi}|^2) \  &\geq& \ \sigma_p^2(\tilde{b},2-q'/2) \frac{\sigma_p^2(\tilde{b},2-q/2)}{\gamma^2}\nonumber \\[2mm]   \ &=&\ \frac{\hbar^2}{4} \frac{q^2}{(3q/2-1)(3q'/2-1)} \ = \ \frac{\hbar^2}{4} f(q)\, .
\label{SM.84.cc}
\end{eqnarray}
Note, that the unknown $\tilde{b}$ parameter completely factored out from the RHS of (\ref{SM.84.cc}). The function $f(q)$ is positive and monotonically increasing for $q\in [1,2]$ with $\max f(q) = 1$.
It is important to stress that $f(q)$ is an universal function of $q$.
So, by using the fact that
\begin{eqnarray}
f(q) \ = \ \left[\frac{2}{(|2/q -1| + 1) (3q/2-1)}\right] \left[\frac{2}{(|2/q' -1| + 1)(3q'/2-1)}\right] \ = \ \phi(q/2)\phi(q'/2)\, ,
\end{eqnarray}
we may define the rescaled entropy power $\tilde{M}_{x}^T = \phi^{-1}(x)M_{x}^T$ and rewrite the EPUR in the form
\begin{eqnarray}
\tilde{M}_{q/2}^T(|{\psi}|^2)\tilde{M}_{q'/2}^T(|\tilde{\psi}|^2) \  \geq  \ \frac{\hbar^2}{4}\, .
\end{eqnarray}
It is this form of the EPUR that is used in the main text (with tilde omitted).

Since Tsallis distribution maximizing Tsallis
entropy ${\mathcal{S}}_{q'/2}^T$ has the non-extensivity parameter $2- q'/2$, and because for $\beta < 0$  the non-extensivity parameter
$2- q'/2 < 1$ for $q' > 2$ we have (as expected) that $q < 2$. The lower bound for $q$ is fixed by the Beckner--Babenko inequality  to be $\geq 1$.

%
%
%

\section*{G. ~$(\Delta \bm{p})^2_{\psi}$ and ultra-relativistic equipartition theorem  \label{rept}}

Here we provide a simple evaluation of $(\Delta \bm{p})^2_{\psi}$ the ultra-relativistic limit by employing equipartition theorem. Let us first use the general equipartition relation~\cite{Huang}
\begin{eqnarray}
\left \langle p_i \frac{\partial H}{\partial p_i} \right\rangle \ = \ k_B T\, .
\end{eqnarray}
For a relativistic particle (no modification of the dispersion relation at high energies is assumed) this implies
\begin{eqnarray}
\left \langle \bm{p}\cdot \frac{\partial H} {\partial \bm{p}}
\right \rangle  \ = \ \sum_{i=1}^3 \left\langle  p_i \frac{cp_i}{\sqrt{\bm{p} + m^2 c^2}}\right\rangle \ = \ \left\langle \frac{\bm{p}^2 c^2}{p_0} \right\rangle\, ,
\end{eqnarray}
which in the ultra-relativistic limit gives $\langle |\bm{p}| \rangle c = 3k_B T$.

In order to find $ \langle \bm{p}^2 \rangle$ we can employ the fact that
the usual derivation of the equipartition theorem also directly  gives the identity
\begin{eqnarray}
\left \langle p_i \frac{\partial H^2}{\partial p_i}\right \rangle \ = \ 2 k_B T \langle H \rangle \ + \ 2 k_B T \left \langle p_i \frac{\partial H}{\partial p_i}\right\rangle\, .
\end{eqnarray}
While the LHS gives
\begin{eqnarray}
\sum_{i=1}^3 \left \langle p_i \frac{\partial H^2}{\partial p_i}\right \rangle
\ = \ 2 c^2 \langle \bm{p}^2 \rangle \ = \ 2 c^2 (\Delta \bm{p})^2 \, ,
\end{eqnarray}
the RHS implies in the ultra-relativistic limit
\begin{eqnarray}
6 k_B T\langle H \rangle \ + \ 6 (k_B T )^2 \ = \ 6 k_B T \langle |\bm{p}| \rangle c \ + \ 6 (k_B T )^2  \ = \ 24 (k_B T )^2\, .
\end{eqnarray}
Consequently, we get in the ultra-relativistic limit that
\begin{eqnarray}
c^2 \langle \bm{p}^2 \rangle \ = \ c^2 (\Delta \bm{p})^2 \ = \ 12 (k_B T )^2\, .
\end{eqnarray}
Since inflaton's rest mass $\simeq 10^{12}-10^{13}$GeV, cf. e.g. Ref.~\cite{McDonald}, the inflaton is during the late inflation/reheating epoch (i.e., $\simeq 10^{15}-10^{16}$GeV) in ultra-relativistic regime. If we now employ the fact that in semi-classical regime thermal and quantum fluctuations should be of the same order we get for inflaton $(\Delta \bm{p})^2_{\psi} \simeq 12 (k_B T )^2$.